\documentstyle[prd,aps]{revtex}
\begin{document}
\input epsf
\draft
\twocolumn[\hsize\textwidth\columnwidth\hsize\csname
@twocolumnfalse\endcsname
\preprint{SU-ITP-98-05, gr-qc/9802038}
\title{Quantum Creation of  an Open Inflationary Universe}
\author{Andrei Linde}
\address{Department of Physics, Stanford University, Stanford, CA
94305, USA}
\date{17 February 1998}
\maketitle
\begin{abstract}
We discuss a dramatic difference between the description of the quantum
creation of an open universe using the Hartle-Hawking wave function and the
tunneling wave function. Recently Hawking and Turok have found that the
Hartle-Hawking wave function leads to a universe with $\Omega = 0.01$, which is
much smaller that the observed value of $\Omega$. Galaxies in such a universe
would be $10^{10^8}$ light years away from each other, so the universe would be
practically structureless.  We argue  that the Hartle-Hawking wave
function does not  describe the probability of creation of the universe. If one
uses the tunneling wave function for the description of creation of the
universe, then in most inflationary models the universe should have $\Omega =
1$, which agrees with the standard expectation that inflation makes the
universe flat. The same result can be obtained in the theory of a
self-reproducing inflationary universe, independently of the issue of initial
conditions. However, there exist some models where $\Omega$ may
take any value, from $\Omega > 1$ to $\Omega \ll 1$.
\end{abstract}
\pacs{PACS: 98.80.Cq  \hskip 3.8cm SU-ITP-98-05
\hskip 3.8cm  gr-qc/9802038}
\vskip2pc]

\section {Introduction}

Until very recently it was believed that the universe after inflation must
become extremely flat, with $\Omega
= 1 \pm 10^{-4}$. This
implied that if observational data
show that $\Omega$ differs from $1$ by more than a fraction of a percent,
inflationary theory should
be ruled out. Of course, it was always possible to
make inflation short, and $\Omega$ different from 1 by fine tuning the
parameters, but in this case the problems of homogeneity and isotropy of the
observable part of the universe would remain unsolved.

Fortunately, this problem was solved recently.
The main idea is to use the well
known fact that the region of
space created in the process of
quantum tunneling tends to have
spherically
symmetric shape, and homogeneous interior, if the tunneling probability is
suppressed strongly enough. Then such bubbles of the
new phase tend to   expand
in a spherically symmetric fashion. Thus, if one could associate the
whole visible part of the universe with an interior of one such region, one
would solve the homogeneity and isotropy problems, and then all other problems
would be solved
by the subsequent relatively short stage of inflation.

For a closed universe the realization of this program could be relatively
straightforward \cite{Open}. One should consider the process of quantum
creation
of a closed inflationary universe from ``nothing.'' If the probability of such
a
process is exponentially suppressed (and this is indeed the case if inflation
is
possible only at the energy density much smaller than the Planck density
\cite{Creation,ZelStar,Rubakov,Vilenkin}), then the universe created that way
will be rather homogeneous from the very beginning. Typically it will grow
exponentially large, and $\Omega$ will gradually approach the flat-space limit
$\Omega = 1$. However, there exist many inflationary models where the total
duration of inflation cannot be longer than $60$ to $70$ e-foldings. In such
models the present value of $\Omega$ can be noticeably higher than~$1$. These
models have several potential drawbacks which will be discussed in the last
section of this paper, but nevertheless creation of a closed inflationary
universe, at least in principle, does not seem impossible.

The situation with an open universe is much more complicated. Indeed, an open
universe is infinite, and it may seem impossible to create an infinite universe
by a tunneling process. However, this is not the case: according to Coleman and De Luccia, any bubble formed in the
process of the false vacuum decay looks from inside like an infinite open
universe \cite{CL,Gott}.
 If this universe continues inflating
inside the bubble then we obtain an open inflationary universe.

Until a short while ago
it was not quite clear whether it is possible to
realize the one-bubble open universe scenario in a natural way. An important
step in this direction was
made when the first semi-realistic models of open inflation were proposed
\cite{BGT}. These models were based on investigation of chaotic inflation and
tunneling in
the theories of one scalar field $\phi$. However, as was shown in \cite{Open},
in the simplest versions of such theories with potentials of the type of
${m^2\over 2} \phi^2-{\delta\over 3} \phi^3 + { \lambda\over 4} \phi^4$ the
tunneling does not occur
by bubble
formation, but by jumping onto the top of the potential barrier described by
the
Hawking-Moss instanton \cite{HM}. This instanton was originally interpreted as
describing homogeneous tunneling, but later it was found that this is not the
case \cite{Star,Gonch,Star2,book,Lab}. This process leads to the
formation of inhomogeneous
domains of a
new phase, and the whole scenario fails. 

This problem   is in fact rather general, it appears  not only in the models with the potential ${m^2\over 2} \phi^2-{\delta\over 3} \phi^3 + { \lambda\over 4} \phi^4$.    Indeed,  Coleman-De Luccia instantons by their construction must be smaller than the size of the Euclidean continuation of de Sitter space $H^{-1}$. Meanwhile, the typical size of a bubble is of the same order as the inverse mass of the field $\phi$, which can be estimated as  $1/\sqrt{V''(\phi)}$. This implies that  these instantons can exist only if $V''(\phi) \gg H^2$ inside the bubble. This condition is incompatible with the assumption of Ref. \cite{BGT} that  inflation continues after the tunneling, which would  require  that $V''(\phi) \ll H^2$ inside the bubble.

In order to resolve this problem one is
forced  to ``bend'' the  effective potentials in a rather specific way.  The
potential must be very flat everywhere except at
one place where it should have a
very deep minimum and a sharp maximum. In addition, one should consider models where inflation inside the bubble begins not immediately after the tunneling, but much later.  These requirements   make  the corresponding
models of open inflation not only fine-tuned but also very complicated. No realistic versions of open inflation models of this type   have been invented so far.

Fortunately, the same goal can be  achieved if one considers models of two
scalar fields \cite{Open}. The presence of two scalar fields allows one to
obtain the required bending of the inflaton potential by simply changing the
definition of the inflaton field in the process of inflation. The tunneling   occurs with respect to a heavy field $\sigma$ with a steep barrier in its
potential, while after the tunneling the role of the inflaton is played by a
light field $\phi$, rolling along
a flat direction ``orthogonal'' to the direction of
quantum tunneling. Inflationary models of this type are quite simple, yet they
have many interesting features. In these models the universe consists of
infinitely many expanding bubbles immersed into an
exponentially expanding false
vacuum state. Each of these bubbles on the
inside looks like an infinitely large open
universe, but the values of $\Omega$ in these universes may take any value from
$1$ to $0$.

Many versions of these two-field models have been considered in the recent
literature, see
e.g. \cite{Open,LidOc,HOpen}. Some of them did not survive comparison with
the
observational data, some of them are very fine-tuned, but in any case one can
no
longer claim that inflation and open universe are incompatible. The simplest
open inflationary model of this type describes two scalar fields with the
effective potential
\begin{equation}\label{4a}
V(\phi,\sigma) = {g^2\over 2}\phi^2\sigma^2 + V(\sigma) \ ,
\end{equation}
where the effective potential for the field $\sigma$ can be taken, e.g., in the
following form:
 $V(\sigma) = {M^2\over 2} \sigma^2 -{\alpha M
} \sigma^3 + {\lambda\over 4}\sigma^4 +V_0$ \cite{Open}. Here $V_0$ is a
constant which is added to ensure that $V(\phi,\sigma) = 0$ at the absolute
minimum of $V(\phi,\sigma)$. If the initial value of the field $\phi$ is
sufficiently large, then the field $\sigma$ is trapped at $\sigma = 0$. The
field $\phi$ slowly drifts in different directions due to inflationary quantum
fluctuations, and in the regions where it becomes  smaller than certain critical value $\phi_c$, the phase
transition to large $\sigma$ becomes possible. Inside the bubbles of the field
$\sigma$ the field $\phi$ acquires nonvanishing mass squared $g^2\sigma^2$, it
begins to slide
towards $\phi = 0$, and yields the
secondary stage of inflation.
Depending on the initial value of the field $\phi$, this stage may be either
short, creating open universes with small $\Omega$, or long, creating
universes with $\Omega \approx 1$. If the probability of
bubble production
is very small, the vacuum state with $\sigma = 0$ will never completely decay,
and the process of creation of new bubbles will never end. This implies that in
an eternally existing
self-reproducing universe based on this scenario there will be infinitely many
universes containing any particular value of $\Omega$, from $\Omega = 0$ to
$\Omega = 1$.  Moreover, the effective value of $\Omega$ in this scenario may vary even within each of the bubbles \cite{BGM}.

An intriguing possibility which will be discussed in this paper is quantum creation of an open universe from nothing. Until very recently such a process seemed impossible.
Indeed, in accordance with the investigation of inflationary universe creation
performed in \cite{Creation,ZelStar,Rubakov,Vilenkin}, the probability of
quantum creation of an inflationary universe is
expected to be suppressed by $e^{-|2S|}$, where $S$ is the value of Euclidean
action on the trajectory describing the universe creation. For a closed
universe
with vacuum energy $V(\phi,\sigma)$ one has
\begin{equation}\label{TUNN}
P \sim e^{- 2|S|} = \exp \left(-{3 M_p^4\over 8 V(\phi,\sigma)}\right) \ .
\end{equation}

One could expect that the action $S$ on an instanton describing the
creation of
an infinitely large  open universe must be infinitely large. Hence
one would not
expect that an open universe can be created unless it is topologically
nontrivial and compact \cite{opencr}. However, this problem disappears in the
new class of open universe models considered above. The probability of
quantum creation of a closed
inflationary universe is finite. After its creation it inflates and becomes
flat
and practically infinite. In the scenario described above, it unceasingly
produces more and more bubbles, each of which represents a new
infinite open universe. Thus, in this scenario one does not
encounter any problems in creating an open universe from nothing. In fact one
does not create
a single open universe but infinitely many of them, with different
values of $\Omega$ in each of the universes \cite{Erice}.

Recently the possibility of
quantum creation of an open universe was pursued
even further in a paper by Hawking and Turok \cite{HT}. They argued that an
open
universe can be created from nothing even without passing through
an intermediate stage
of
false vacuum inflation and subsequent tunneling. According to \cite{HT}, this regime
is possible in the theories of a single field $\phi$ with the
simplest potentials of the chaotic inflation type \cite{Chaotic}. This would be a
very interesting
and encouraging development. However, Hawking and Turok
used the Hartle-Hawking wave function of the universe
\cite{HH} to describe the probability of creation of
an open universe.
As a result, they
 experienced severe problems usually
associated with
the description of the universe creation in the context of the Hartle-Hawking
approach.
Typical universes produced by the process described in \cite{HT} tend to be not
only
open, but entirely empty, $\Omega
\to 0$. The only way to avoid this disastrous conclusion
is to use anthropic principle and argue that we live in a universe with small
$\Omega$ simply because we cannot live in the universe with $\Omega = 0$.
But even this  does not help much. Estimates made in \cite{HT} show that   the
maximum of probability to live in
an open universe is sharply
peaked at $\Omega = 0.01$, which does not agree with the observational data.

In this paper we will show that this result is practically model-independent,
and it appears solely due to the use of the
Hartle-Hawking wave function. This wave function gives the probability of the
universe creation of a very peculiar form,
\begin{equation}\label{HH}
P \sim e^{- 2S } = \exp \left({3 M_p^4\over 8 V(\phi)}\right) \ ,
\end{equation}
which strongly disfavors inflation of any kind and suggests that it is much
easier to create an infinite Minkowski space rather that a Planckian size
closed universe. The difference between Eqs.
(\ref{TUNN}) and (\ref{HH}) appears due to the ``wrong'' sign of the
gravitational action of the instanton describing creation of de Sitter
universe,
$S =- {3 M_p^4\over 16 V(\phi)}$.

As it was argued in \cite{Creation,book,LLM}, the Hartle-Hawking wave
function does not describe the probability of the universe creation. Rather,
it describes the probability of quantum fluctuations in a universe which
has already
been born. In particular, the probability distribution (\ref{HH}) implies that the universe in its ground state lives near the minimum of the effective potential, and the probability of its deviations from this state is exponentially small. 

Meanwhile the essence of inflationary theory is that initially the universe could be very far from the minimum of $V(\phi)$. It takes a lot of time for the field to roll to this minimum, and during this time the universe becomes exponentially large. Thus, in our opinion, the tunneling wave function makes an attempt to describe  creation of an inflationary universe, inflationary theory tells us how the universe approached the minimum of $V(\phi)$, whereas  the Hartle-Hawking wave function describes properties of the universe after it reaches its ground state, in case if such a ground state  exists.  

To clarify this issue, in Sect. II of this paper we will recall
the history
of the debate related to the choice of the
Hartle-Hawking versus the tunneling wave function.
In Sect. III we will analyse this issue again, using the
stochastic approach to
inflation. This
will allow one to have a better understanding of different approaches to the
calculation of the most probable value of $\Omega$ in the context of quantum
cosmology.

Then in Sect. IV
we will discuss the properties of the Hawking-Turok instanton and the
probability of an open universe creation. We will explain the origin of the
result $\Omega \sim 0.01$, and show that  this number   practically
does not depend on the choice of a particular inflationary
model. We will also argue that if one applies the Hartle-Hawking approach to
the creation of the universe, then this  result will endanger all previous
versions  of the open universe scenario \cite{Gott,BGT,Open,LidOc,HOpen}.

We will, however, show
that if one uses the tunneling wave function of the
universe for the description of creation of the universe
\cite{Creation,ZelStar,Rubakov,Vilenkin}, a
typical universe to be created in the simplest versions of the chaotic
inflation
scenario with polynomial potentials will have $\Omega = 1$, rather than $\Omega
= 0.01$. This result is in agreement with the usual
expectation  that inflation typically leads to $\Omega = 1$. However, there
exist several  versions of the chaotic inflation scenario discussed in
\cite{Open}, and one recently proposed version of the hybrid inflation
scenario in  supergravity \cite{Riotto}, where the typical duration of
inflation is very
small. In such models the most probable value of $\Omega$ can take any value
between $1$ and $0$ depending on the parameters of the model, without any need
to appeal to the anthropic principle. These models, however, have serious  problems
of their own, which require further investigation.   If the mechanism which we
will discuss  is successful,
we will have a new class of open inflationary
models. 

In Sect. V we will describe some problems with the more recent  proposal of Hawking and Turok related to the theory of the four form field strength \cite{HT2}.  We will argue that the unfortunate prediction $\Omega = 0.01$ appears in this theory as well.

Independently of the success or failure of the new class of models of open inflation, we will show that the use of the tunneling wave function for the description of the universe creation  preserves the validity of the previous models of open inflation, proposed in
\cite{Gott,BGT,Open,LidOc,HOpen}. Moreover, we will argue that the  models of open inflation proposed in \cite{Gott,BGT,Open,LidOc,HOpen} remain valid independently
of the choice of the wave function describing initial conditions if one  takes into account the possibility of eternal inflation in these models.

\section{Wave function of the Universe}

\subsection{Why do we need quantum cosmology?}

The investigation of the wave function of the universe goes back to the
fundamental
papers by Wheeler and DeWitt \cite{DeWitt}. However, for a long time it seemed
almost meaningless to apply the notion of the wave function to the universe
itself, since the universe is not a microscopic object. Only with the
development of inflationary cosmology it became clear that the whole universe
could appear from a tiny part of space as small as the Planck length $M_p^{-1}$
(at least in the chaotic inflation scenario \cite{Chaotic}). Such a tiny region
of space can appear as a result of quantum fluctuations of metric, which should
be studied in the context of quantum cosmology. Later it was found that the
global structure of the universe in the chaotic inflation scenario is
determined not by classical physics, but by quantum processes \cite{book}.

Unfortunately, quantum cosmology is not a well developed science. This theory
is
based on the Wheeler-DeWitt equation, which is the Schr\"{o}dinger equation for
the wave function of the universe. This equation has many solutions, and at the
present time the best method to specify preferable solutions of this equation,
as well as to interpret them, is based on the Euclidean approach to quantum
gravity. This method is very powerful, but some of its applications are not
well
justified. In some cases this method may give incorrect answers, but rather
paradoxically sometimes these answers appear to be correct when applied
to some other questions. Therefore it becomes necessary not only to solve the
problem in the Euclidean approach, but also to check, using one's best
judgement, whether the solution is related to the original problem or to
something else. An alternative approach is based on the use of stochastic
methods in inflationary cosmology \cite{Star,Gonch,Star2,book,LLM}. These
methods allow  one to understand such
effects as the
creation of inflationary density perturbations, the theory of
tunneling, and even the theory of self-reproduction of inflationary universe.
Both Euclidean approach and stochastic approach to inflation have their
limitations, and it is important to understand them.

\subsection{Hawking-Moss tunneling}

Before discussing quantum creation of the universe, let us pause a little and
study the problem of tunneling between two local minima of the effective
potential $V(\phi)$ in inflationary cosmology. As we will see, this subject is
 closely related to the issue of quantum creation of the universe.

Consider a theory with an effective potential $V(\phi)$ which has a local
minimum at $\phi_0$, a global minimum at $\phi_*$ and a barrier separating
these two minima, with the top of the barrier positioned at $\phi = \phi_1$.
One of the first works on inflationary cosmology was the paper by Hawking and
Moss \cite{HM} where they studied a possibility of tunneling from $\phi_0$ to
$\phi_*$ in the new inflationary universe scenario.

They have written   equations of motion for the scalar field in an
Euclidean space with the metric
\begin{equation}\label{metric}
ds^2 =d\tau^2 +a^2(\tau)(d \psi^2+{\rm sin}^2 \psi d \Omega_2^2) \ .
\end{equation}

The field $\phi$ and the radius $a$ obey
the field equations
\begin{equation}\label{equations}
\phi''+3{a'\over a}\phi'=V_{,\phi},~~~~~ a''= -{8\pi G\over 3} a (
\phi'^2 +V) \ ,
 \end{equation}
where primes denote derivatives with respect to $\tau$.

If the potential has an extremum at some particular value of the field $\phi$,
then the equation for the field $\phi$ is solved trivially by the field staying
at this extremum. Then the equation for $a(\tau)$ has a simple solution
$a(\tau)=H^{-1} {\sin}
(H\tau)$, with $H^2= 8 \pi G V(\phi)/3 = 8 \pi   V(\phi)/3M_p^2$. This solution
describes a sphere $S^4$, the Euclidean version of de Sitter space. In this
description $\tau$ plays the role of Euclidean time, and $a(\tau)$ the role of
the scale factor. One can try to interpret one
half of this sphere as an
instanton. The action on this instanton is negative,
\begin{equation}\label{action}
 S = \int d^4 x \sqrt{-g}\left(-{RM_p^2\over{16\pi }}+V(\phi)\right) = -
{3M_p^4\over 16 V(\phi)}  \ .
\end{equation}

It was argued in \cite{HM} that the probability of tunneling from $\phi_0$ to
the true vacuum $\phi_*$ is given by
\begin{equation}\label{HMinst}
P \sim \exp\left({3M_p^4\over 8V(\phi_1)}\right) \  \exp\left(-{3M_p^4\over 8
V(\phi_0)}\right)
{}~.
\end{equation}
The probability of tunneling, as usual,  is suppressed by $e^{-2S}$ (or by
$e^{-S}$ if by $S$ we mean the result of integration over the whole sphere,
$-{3M_p^4\over 8V(\phi)}$). This is the standard result of  the Euclidean
theory of tunneling. Everything else about this result was rather mysterious.

First of all, instantons typically interpolate between the initial vacuum state
and the final state. Here, however, the scalar field on the instanton solution
was exactly constant. So why do we think that they describe tunneling from
$\phi_0$ if $\phi_0$ never appears in the instanton solution?

\begin{figure}[Fig0111]
 \hskip 1.5cm
\leavevmode\epsfysize=7cm \epsfbox{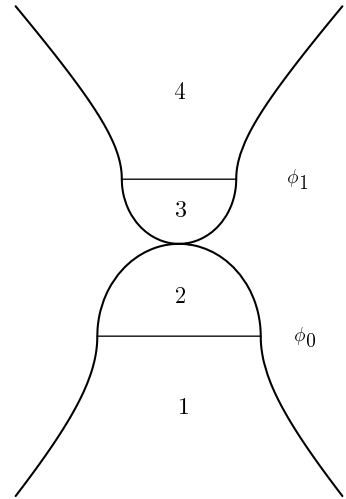}

\

\caption[Fig1]{\label{HawkMoss} A possible interpretation of the Hawking-Moss
tunneling from $\phi_0$ to $\phi_1$.}

\end{figure}

A possible answer to this question can be given as follows. One can choose the
coordinate system where inflationary universe looks as a  closed de Sitter
space near the point of a maximal contraction, where its size becomes
$H^{-1}(\phi_0)$, see region 1 in Fig. \ref{HawkMoss}. Classically, such a
universe at that moment  begins expanding with the same value of the Hubble
constant as before. However, since the total size of the universe at that
moment is finite, it may also jump quantum mechanically to a state with a
different value of the field $\phi$ corresponding to a different extremum of
the effective potential. One can represent this process by gluing two de Sitter
instantons corresponding to  two different values of the scalar field $\phi$
($\phi_0$ in the region  2, and $\phi_1$  in the region 3 in Fig.
\ref{HawkMoss}), and by making analytical continuation to the Lorentzian
regions  1 and 4.

This seems to be a plausible interpretation of the Hawking-Moss tunneling (see also \cite{CB}). But
it certainly does not answer all questions. What will happen
if we have several different local minima and maxima of $V(\phi)$? Why does the
tunneling  go
to the top of the effective potential rather than to the absolute minimum of
the effective potential, or to some other local maximum? Finally, if the
instanton
describes an exactly homogeneous scalar field $\phi$, does it mean that the
tunneling must simultaneously occur everywhere in an exponentially large
inflationary universe? This does not seem plausible, but what else should
we think
about, if the field $\phi$ on the instanton solution is constant?

And indeed, originally it was assumed that the tunneling described by this
instanton must occur simultaneously in the whole universe. Then, in the second
paper on this subject, Hawking and Moss said that their results were widely
misunderstood, and that
this instanton describes tunneling which is homogeneous only
on the scale of horizon $\sim H^{-1}$ \cite{HM2}. But how is it possible to
describe inhomogeneous tunneling by a homogeneous instanton?

A part of the answer was given in Ref. \cite{Gonch}. We have found that if one
deforms a little the Hawking-Moss instanton  to make the field $\phi$ match
$\phi_0$ in some small region of the sphere, we will, strictly speaking, not
get
a solution, but the action on such a configuration can be made almost exactly
coinciding with the Hawking-Moss action. Then such configurations can play the
same role as instantons \cite{Nucl}.

A full understanding of this issue was reached only after the development of
the stochastic approach to inflation \cite{Star,Gonch,Star2,book}. We will
return to this question later.

\subsection{Creation of the universe from nothing}

Now we will discuss  the problem of the universe creation. According
to classical cosmology, the universe appeared from the singularity in a state
of
infinite density. Of course, when the density was greater than the Planck
density $M_p^4$ one could not trust the classical Einstein equations, but in
many cases there is no demonstrated need to study the universe creation using
the methods of quantum theory. For example, in the simplest versions of the
chaotic
inflation scenario \cite{Chaotic}, the process of inflation, at the classical
level, could begin
directly in the initial singularity. However, in certain models, such as the
Starobinsky model \cite{b14} or the new inflationary universe scenario
\cite{New}, inflation cannot start in a state of infinite density. In such
cases
one may speculate about the possibility that inflationary universe appears due
to quantum tunneling ``from nothing.''

The first idea how one can describe creation of an inflationary universe ``from
nothing'' was given
in 1981 by Zeldovich \cite{Zeld} in application to the
Starobinsky model \cite{b14}. His idea was qualitatively correct, but he did
not
propose any quantitative description of this process. A very important step in
this direction was made in 1982 by Vilenkin \cite{NothVil}. He suggested to
calculate the Euclidean action on de Sitter space with the energy density
$V(\phi)$, which coincides with the Hawking-Moss instanton with the action
$S   = -{3M_p^4\over 16 V(\phi)}$. However, as we have seen, this instanton by
itself does not tell us where the tunneling comes from. Vilenkin suggested to
interpret this instanton   as   the tunneling trajectory describing
creation of the universe with the scale factor $a = H^{-1} = \sqrt{3 M_{\rm
P}^2\over 8\pi V}$ from the state with $a = 0$. This would imply that the
probability of quantum creation of the universe is given by
\begin{equation}
{P} \propto \exp (-2S) = \exp \left({3 M_p^4 \over 8 V(\phi)}\right).
\label{Vil1}
\end{equation}
A year later this result received strong support when Hartle and Hawking
reproduced it by a different though closely related method \cite{HH}. They
argued that the wave function of the ``ground state'' of the universe with a
scale factor $a$ filled with a scalar field $\phi$ in the semiclassical
approximation is given by
\begin{equation}\label{E31}
\Psi_0(a,\phi)\sim \exp\left(-S(a,\phi)\right)\ .
\end{equation}
Here $S(a,\phi)$ is the Euclidean action corresponding to the Euclidean
solutions of the Lagrange equation for $a(\tau)$ and $\phi(\tau)$ with the
boundary conditions $a(0)=a, \phi(0)=\phi$. The reason for choosing this
particular wave function was explained as follows. Let us consider the Green's
function of a particle which moves from the point $(0,t')$ to the point ${\bf
x},t$:
\begin{eqnarray}\label{E32}
<{\bf x},t|0, t'>
 &=& \sum_n \Psi_n ({\bf x})\Psi_n(0)
\exp\left(iE_{n}(t-t')\right) \nonumber \\
 &=& \int d{\bf x}(t) \exp\left(iS({\bf x}(t))\right)\ ,
\end{eqnarray}
where $\Psi_n$ is a complete set of energy eigenstates corresponding to the
energies $E_n\geq 0$.

To obtain an expression for the ground-state wave function $\Psi_0({\bf x})$,
one should make a rotation
           $t \rightarrow -i\tau$ and take the limit as
$\tau \rightarrow -\infty$\@. In the summation (\ref{E32}) only the term $n=0$
with the lowest eigenvalue $E_0 = 0$ survives, and the integral transforms into
$\int dx(\tau)\exp(-S({\bf x}(\tau)))$. This yields, in the semiclassical
approximation,
\begin{equation}\label{E31aa}
\Psi_0(x)\sim \exp\left(-S({\bf x})\right)\ ,
\end{equation}
where the action is taken on the classical trajectory bringing the particle to
the point ${\bf x}$. Hartle and Hawking have argued that the
generalization of this result to the case of interest   would yield
(\ref{E31}).

The method described above is very powerful. For example, it provides the
simplest way to find the wave function of the ground state of the harmonic
oscillator in quantum mechanics. However, this wave function simply describes
the probability of deviations of the harmonic oscillator from its equilibrium.
It certainly does not describe quantum creation of a harmonic oscillator.
Similarly, if one applies this method to the hydrogen atom, one can obtain the
wave function of an electron in the  state with the lowest energy. Again, this
result has no relation to the probability of creation of an electron from
nothing.

The gravitational action involved in (\ref{E31}) is the same action as before,
corresponding to one half of the Euclidean section
$S_4$ of de Sitter space with $a(\tau)
= H^{-1}(\phi)\cos H\tau$ ($0\leq\tau\leq H^{-1}$). One can represent it in the
following form:
\begin{eqnarray}\label{E33}
S(a, \phi) &=& - \frac{3\pi M_p^2}{4}
    \int d\eta\Bigl[\Bigl(\frac{da}{d\eta}\Bigr)^2 - a^2 +
\frac{8\pi V}{3M_p^2}a^4\Bigr] \nonumber \\
   &=& - \frac{3M_p^4}{16 V(\phi)}\ .
\end{eqnarray}
Here $\eta$ is the conformal time, $\eta = \int {d\tau\over a(\tau)}$.
Therefore, according to \cite{HH},
\begin{equation}\label{E34}
\Psi_0(a,\phi)\sim \exp{\Bigl(-S(a,\phi)\Bigr)} \sim
\exp\left(\frac{3 M_p^4
}{16V(\phi)}\right) .
\end{equation}
By taking a square of this wave function one again obtains eq. (\ref{Vil1}).
The
corresponding expression has a very sharp maximum as $V(\phi) \rightarrow 0$.
This could suggest  that the probability of finding the universe in a state
with a
large field $\phi$ and having a long stage of inflation should be strongly
suppressed. But is it a correct interpretation of the Hartle-Hawking wave
function? Just like in the examples with the harmonic oscillator and the
hydrogen atom mentioned above, nothing in the `derivation' of the
Hartle-Hawking wave function tells that it describes creation of the universe
from nothing. The simplest way to interpret the Hartle-Hawking wave function in
application to de Sitter space is as follows. At the classical level, de Sitter
space has a definite speed of expansion, definite size of its throat $H^{-1}$,
etc. At the quantum level, de Sitter ``trajectory'' becomes wider because of
quantum fluctuations. The Hartle-Hawking wave function of de Sitter space
describes the probability of deviations of  metric of de Sitter space   from
its classical expectation value, which may occur due to the process shown in
Fig. \ref{HawkMoss}. This is very much different from the probability of
spontaneous creation of the universe.

In fact, Eq. (\ref{Vil1}) from the very beginning did not seem to apply to the
probability of creation of the universe. The total energy of matter in a closed
de
Sitter space with $a(t) = H^{-1}\cosh Ht$ is greater than its minimal volume
$\sim H^{-3}$ multiplied by $V(\phi)$, which gives the total energy of the
universe $E {\
\lower-1.2pt\vbox{\hbox{\rlap{$>$}\lower5pt\vbox{\hbox{$\sim$}}}}\
} M_p^3/\sqrt V$. Thus the minimal value of the total energy of matter
contained in a closed de Sitter universe {\it grows} when $V$ decreases. For
example, in order to create the universe at the Planck density $V
\sim M_p^4$ one needs no more than the Planckian
energy $M_p
\sim 10^{-5}$ g. For the universe to appear at the GUT energy density  $V
\sim M_X^4$ one needs to create from nothing the universe with the total energy
of
matter of the order of $M_{\rm Schwarzenegger} \sim 10^2$ kg, which is
obviously
much more difficult. Meanwhile, if one makes an attempt to use the
Hartle-Hawking wave function for the description of the creation of the
universe (which, as we believe, does not follow from its derivation), then eq.
(\ref{Vil1}) suggests that it should be
much
easier to create a huge universe with  enormously large total
mass rather than a small universe with Planckian mass. This seems very
suspicious.

From uncertainty relations one can expect that the probability of a process
of
universe formation is not exponentially suppressed if it occurs within a time
$\Delta t < E^{-1}$. This is quite possible if the effective potential is of
the
order of $M_p^4$ and $E
\sim M_p^3/\sqrt V \sim M_p$. In such a case one may envisage the process of
quantum creation of a universe of mass $M_p$ within the Planck time $M_p^{-1}$.
However, the universe
of mass $E \gg M_p$ (which is the case for $V \ll M_p^4$) can be created only
if the corresponding process lasts
much shorter than the Planck time $M_p^{-1}$, which is hardly possible.

Another way to look at it is to calculate the total entropy ${\rm \bf S}$ of de
Sitter space at the moment of its creation. It is equal to one quarter of the
horizon area of de Sitter space (in Planck units), which gives ${\rm \bf S} =
{3
M_p^4\over 8 V(\phi)}$. (Note its relation to the Euclidean action on the full
de Sitter sphere ${  S} = -{3
M_p^4\over 8 V(\phi)}$.) It seems natural to expect that the probability of
emergence of a complicated object of large entropy must be suppressed by a
factor of $\exp({-\rm \bf  S}) = \exp(-{3 M_p^4\over 8 V(\phi)})$, which again
brings us
to the equation (\ref{TUNN}), see \cite{KLB}.
Meanwhile the use of the
Hartle-Hawking wave
function for the description of creation of the universe
would indicate that it is
much more probable to create a very large universe with a huge entropy rather
than a small universe with entropy $O(1)$.

To avoid misunderstandings,  one should note, that the probability of
fluctuations in a thermodynamical system is always {\it suppressed} by the
factor $e^{\rm \bf \Delta S}$, where ${\rm \bf \Delta S}$ is the change of
entropy between two different states of the system  \cite{HTnew}. As we will
see, this is exactly what happens during the tunneling between two different
states of de Sitter space with two different values of $V(\phi)$. This is in
perfect agreement with the prediction of the Hartle-Hawking wave function if
one applies it not to the creation of the universe but to the probability of
its change.   However, now we are not talking about   the probability of change
of the state of the  system, but about a possibility of creation of the whole
system together with a lot of information stored in it from nothing. We are not
going to insist that this process is possible. In fact in chaotic inflation
scenario this assumption is not necessary because the universe formally can
inflate even in a   state with indefinitely large density, so there is no need
for any tunneling to take place. However, if creation from nothing is possible
at all, then the tunneling wave function suggests that this process should be
as unintrusive as possible, whereas the Hartle-Hawking approach implies that
the greater  the change, the easier it occurs. I leave it for the reader to
decide
whether this
looks plausible.

One may wonder why the Hartle-Hawking wave function leads to  rather
counterintuitive predictions when applied to the probability of creation of the
universe? There is one obvious place where the derivation
(or interpretation) of eq.
(\ref{Vil1}) could go wrong. The effective Lagrangian of the scale factor $a$
in
(\ref{E33})   has a wrong overall sign. Solutions of the Lagrange equations do
not know anything about
the sign of the Lagrangian, so we may simply change the sign before studying
the
tunneling. Only after switching the sign of the Lagrangian of the scale factor
in (\ref{E33}) and representing the theory in a conventional form can we
consider tunneling of the scale factor. But after changing the sign of the
action, one obtains a different expression for the probability
of quantum creation of the universe:
\begin{equation}
{P} \propto \exp (-2|S|) = \exp \left(-{3 M_p^4\over 8 V(\phi)}\right).
\label{E366}
\end{equation}
This equation predicts that a typical initial value of the field $\phi$ is
given
by $V(\phi)\sim M_p^4$ (if one does not speculate about the possibility that
$V(\phi) \gg M_p^4)$, which leads to a very long stage of inflation.

Originally I obtained this result by the method described above. However,
because of the ambiguity of the notion of tunneling from the state $a = 0$, one
may try to look at the same subject from a different perspective, and reexamine
the derivation of the Hartle-Hawking wave function. In this case the problem of
the wrong sign of the Lagrangian appears again, though in a somewhat different
form.
  Indeed,   the
total energy of a closed universe is zero, being a sum of the positive energy
of
matter and the negative energy of the scale factor $a$. Thus, the energy $E_n$
of the scale factor is negative. If one makes the same Euclidean rotation as in
Eq. (\ref{E32}), the contributions of all states with $n >1$ will be greater
than the contribution of the state with the lowest absolute value of energy, so
such a rotation would not allow one to extract the wave function $\Psi_0$ as we
did before. This is a simple mathematical fact, which means that the main argument used in \cite{HH} to justify their prescription of quantization of the scale factor fails.

In order to suppress terms with large
negative $E_n$ and to obtain $\Psi_0$ from (\ref{E32}) one should rotate $t$
not
to $-i\tau$, but to $+i\tau$. This gives \cite{Creation}
\begin{equation}\label{E35a}
\Psi_0(a,\phi) \sim \exp\Bigl(-|S(a,\phi)|\Bigr) \sim\exp
\left(- \frac{3 M_p^4}{16V(\phi)}\right) ,
\end{equation}
and
\begin{equation}\label{E36}
P(\phi) \sim|\Psi_0(a,\phi)|^2 \sim\exp
\left(- \frac{3 M_p^4}{8V(\phi)}\right)   .
\end{equation}
Later this equation was also derived by
 Zeldovich and Starobinsky \cite{ZelStar},
 Rubakov \cite{Rubakov}, and Vilenkin \cite{Vilenkin} using the methods similar
to the first
method mentioned above (switching the sign of the Lagrangian). The
corresponding wave function (\ref{E35a}) was called
``the tunneling wave function.'' This wave function\footnote{In fact, the two different ``derivations'' of this wave function  described above lead to two slightly different wave functions \cite{vilrecent}. However, since the difference between these two versions of the tunneling wave function is exponentially small, we will neglect it in this paper.} is dramatically different from the
Hartle-Hawking wave function \cite{HH}, as well as from the Vilenkin's wave
function proposed few years earlier \cite{NothVil}.

An obvious objection against this result is that it may be incorrect
to use  different  ways of rotating $t$ for the quantization of the
scale factor and of the scalar field, see e.g. \cite{HTnew}. If one makes the
same rotation for for the matter fields as the rotation which we proposed for
the scale factor, then one may encounter catastrophic particle production and
other equally unpleasant consequences.
On the other hand, as we have seen, if one assumes  without any proof  that it
is enough to make the standard Wick rotation to quantize the scale factor, one
does not obtain the wave function of the ground state $\Psi_0$, and one gets
the counterintuitive result that large universes are created much easier than
the small ones.

 We believe that  the problem here goes far beyond the issue of the Wick
rotation. The idea that a consistent
quantization of an unstable system of matter with positive energy density
coupled to gravity with negative energy density can be accomplished by a proper
choice of a complex contour of integration may be too optimistic. We know, for
example, that despite many attempts to develop a Euclidean formulation of
nonequilibrium quantum statistics or of
the field theory in a nonstationary
background, such a formulation still does not exist. It is quite clear from
(\ref{E32}) that the $t \rightarrow -i\tau$ trick does not give us the ground
state wave function $\Psi_0$ if the
spectrum
$E_n$ is not bounded from below. Absence of equilibrium, of any simple
stationary ground state, seems to be a typical situation in quantum cosmology.
A
closely related instability is the basis of inflationary cosmology, where
exponentially growing total energy of the scalar field appears as a result of
pumping energy  from the gravitational filed, whereas the total energy of
matter plus gravitational field remains zero.

Fortunately, in certain limiting cases this issue can be resolved in a
relatively
simple way. For example, at present the scale factor $a$ is very big and it
changes
very slowly, so one can consider it as a classical background, and quantize
only the usual matter fields with positive energy. In this case one should use
the standard Wick rotation $t
\rightarrow
-i\tau$\@. On the other hand, in
inflationary universe the evolution of the scalar field is very slow; during
the
typical time intervals $O(H^{-1})$ it behaves essentially as a classical field.
Thus to a good approximation
 one can  describe the process of  creation of an inflationary universe
filled with a homogeneous scalar field by the quantization of the scale factor
$a$
only, and by the rotation $t \rightarrow i\tau$. When using the tunneling wave
function, for example, for the description of particle creation in de Sitter
space, instead of introducing a universal rule for the Wick rotation one should
operate in a more delicate way, treating separately the scale factor and the
particle excitations, see e.g.  \cite{VG}.   

Similarly, one should not use the
Hartle-Hawking wave function for the description of creation of an
inflationary universe, but one can use it for investigation of  fluctuations of
this background. These fluctuations are local, and often they appear simply as a result of quantum fluctuations of matter fields having positive energy. 
In particular, long-wavelength fluctuations of the scalar field $\phi$ in inflationary universe may change local value of energy density $V(\phi)$ inside the domains of a size greater than the size of the event horizon $H^{-1}$. For a comoving observer, such a change   looks like a homogeneous change of the scalar field $\phi$ and of the Hubble constant $H(\phi)$, so he might want to (erroneously) interpret it as a result of quantum fluctuations of the scale factor. These are local perturbations of the homogeneous classical background. These perturbations are produced by fields with positive energy. Therefore in all situations where the inflationary background changes slowly (and in this sense can be considered a ground state of the system) one can use the Hartle-Hawking wave function for investigation of fluctuations of this background.

For
example, Hartle-Hawking wave function can be used for description of black hole formation in a pre-existing de Sitter background \cite{BH}. But this method should not be used for description of quantum creation of de Sitter space with a pair of black holes in it. 

One can also obtain the amplitude of density perturbation in inflationary
universe by a rather complicated method using the Hartle-Hawking wave function
\cite{HHAL}. However, the same results for density perturbations can be obtained by assuming that
inflationary universe was created from nothing in accordance with the tunneling
wave function, and then it expanded and produced perturbations in accordance to
\cite{Pert}. Moreover, as we already mentioned, in chaotic inflation there is
no need to assume that any process of tunneling ever took place in the early
universe. One may simply assume that the universe from the very beginning
expanded classically, and then obtain the same results for the density
perturbations using methods of Ref. \cite{Pert}.

Derivation of equations (\ref{Vil1}), (\ref{E36}) and their interpretation is
far from being rigorous, and therefore even now it remains a subject of
debate. From time to time this issue attracts a lot of attention. For example,
the famous proposal to solve the cosmological constant problem in the context
of the baby universe theory,  which was very popular ten years ago, was based
entirely  on the use of the  wrong  sign of de Sitter action in the
Hartle-Hawking approach to quantum gravity \cite{b66,Coleman}. One of the main
authors of this proposal, Sidney Coleman, emphasized: ``The euclidean
formulation of gravity is not a subject with firm foundations and clear rules
of procedure; indeed it is more like a trackless swamp. I think that I have
threaded my way through it safely, but it is always possible that unknown to
myself I am up to my neck in quicksand and sinking fast'' \cite{Coleman}. After
two years of intensive investigation of this issue it became clear that   the
wrong sign of the Euclidean action can hardly provide a reliable explanation
for the vanishing of the cosmological constant.  Moreover, recent observational
data indicate that the cosmological constant may not vanish after all.

To summarize, the derivation of the Hartle-Hawking wave function is rather
ambiguous. Still, our main objection  with respect to this wave function is
related not to its derivation, but rather to its interpretation. The main
purpose of the paper by Hartle and Hawking \cite{HH} was to find the wave
function describing the least exited, stationary state of the gravitational
system, which would be analogous to the ground state on the harmonic oscillator
or of the hydrogen atom. And indeed it  gives a nice description of  quantum
fluctuations near  de Sitter background, which in a certain sense is
stationary. (There is a coordinate system where de Sitter space is static.)
In such a situation one can consider matter fluctuations, and then find fluctuations of the scale factor induced by the fluctuations of matter. Then the problem of negative energy of the scale factor does not arise, and one can use the Hartle-Hawking wave function to study fluctuations in/of the pre-existing background.
However, we do not see anything in the ``derivation'' of the Hartle-Hawking wave function which would indicate that it
can be used for investigation of the probability of quantum creation of the
universe. 

The tunneling wave function also has certain limitations, but it  seems
to have a better chance to describe the process of quantum creation of the universe. In the subsequent discussion an exact form of this wave function will not be important for us. The only property of this wave function which we are going to use is that   quantum creation of the universe should not be strongly suppressed if it can be achieved by fluctuations of metric on the Planck scale $M_p^{-1}$   at the Planck density $M_p^4$.

Since the debate concerning the wave function of the universe continues for the
 last 15 years, it may be useful to look at it from a somewhat different
perspective, which does not involve discussion of   ambiguities  of the
Euclidean quantum gravity. In the next section we will discuss the stochastic
approach to quantum
cosmology. Within this approach equations (\ref{Vil1}) and (\ref{E36}) can be
derived in a much more clear and rigorous way, but they will have a somewhat
different interpretation.

\section{Wave function of the universe and stochastic approach to inflation}

In this section we will briefly describe the stochastic approach to inflation
\cite{Star,book,LLM}. It is less ambitious, but also much less ambiguous than
the approach based on the investigation of the wave function of the universe.
One of the tools used in this approach is the probability distribution
$P_c(\phi, t|\phi_0)$, which describes the probability of finding the field
$\phi$ at a given point at a time $t$, under the condition that at the time
$t=0$ the field $\phi$ at this point was equal to $\phi_0$. The same function
also describes the probability that the scalar field which at time $t$ was
equal
to $\phi$, at some earlier time $t=0$ was equal to $\phi_0$.

The  probability distribution $P_c$  is  in fact the
probability distribution per unit volume in {\it comoving coordinates} (hence
the index $c$ in $P_c$), which do not change during the expansion of the
universe.
By considering this probability distribution, we neglect the main source of
self-reproduction of inflationary domains, which is the exponential growth of
their volume. Therefore, in addition to $P_c$, we introduced the probability
distribution $P_p(\phi,\phi_0,t)$, which describes the probability to find a
given field configuration in a unit physical volume \cite{b19,book}.

 Consider the simplest model of chaotic inflation based on the
theory of a scalar field $\phi$ minimally coupled to gravity, with the
effective
potential $V(\phi)$. If the classical field $\phi$ is sufficiently homogeneous
in some domain of the universe, then its behavior inside this domain is
governed
by the equation $3H\dot\phi =
-dV/d\phi$, where $H^2 =
\frac{8\pi V(\phi)}{3 M_p^2 }$.

Inflation stretches all initial inhomogeneities. Therefore, if the evolution of
the universe were governed solely by classical equations of motion, we would
end
up with an extremely smooth universe with no primordial fluctuations to
initiate
the growth of galaxies. Fortunately, new density perturbations are generated
during inflation due to quantum effects. The wavelengths of all vacuum
fluctuations of the scalar field $\phi$ grow exponentially in the expanding
universe. When the wavelength of any particular fluctuation becomes greater
than
$H^{-1}$, this fluctuation stops oscillating, and its amplitude freezes at some
nonzero value $\delta\phi (x)$ because of the large friction term
$3H\dot{\phi}$
in the equation of motion of the field $\phi$\@. The amplitude of this
fluctuation then remains almost unchanged for a very long time, whereas its
wavelength grows exponentially. Therefore, the appearance of such a frozen
fluctuation is equivalent to the appearance of a classical field $\delta\phi
(x)$ that does not vanish after averaging over macroscopic intervals of space
and time.

Because the vacuum contains fluctuations of all wavelengths, inflation leads to
the creation of more and more perturbations of the classical field with
wavelengths greater than $H^{-1}$\@. The average amplitude of such
perturbations
generated during a time interval $H^{-1}$ (in which the universe expands by a
factor of e) is given by
\begin{equation}\label{E23}
|\delta\phi(x)| \approx \frac{H}{2\pi}\ .
\end{equation}
The phase of each wave is random.
Therefore, the sum of all waves at a given
point fluctuates and experiences Brownian jumps in all directions in the field
space.

One can describe the stochastic behavior of the inflaton field using diffusion
equations for the probability distribution $P_c(\phi,t|\phi_0)$. The first
equation is called the backward Kolmogorov equation,
\begin{eqnarray} \label{Starb}
\frac{\partial  P_c(\phi,t|\phi_0)}{\partial t} &=&
  \frac{H^{3/2}(\phi_0)}{8\pi^2}\, \frac{\partial
}{\partial\phi_0}
 \Bigl({H^{3/2}(\phi_0)}  \, \frac{\partial P_c(\phi,t|\phi_0)}{\partial\phi_0}
\Bigr)\nonumber \\
 &-&  \frac{V'(\phi_0)}{3H(\phi_0)} \frac{\partial
P_c(\phi,t|\phi_0)}{\partial\phi_0} \ .
\end{eqnarray}
In this equation one considers the value of the field $\phi$ at the time $t$ as
a constant, and finds the time dependence of the probability that this value
was
reached during the time $t$ as a result of diffusion of the scalar field from
different possible initial values $\phi_0 \equiv
\phi(0)$.

The second equation is the adjoint of the first one; it is called the forward
Kolmogorov equation, or the Fokker-Planck equation \cite{Star},
\begin{eqnarray}\label{E3711}
\frac{\partial P_c(\phi,t|\phi_0)}{\partial t} &=&
   \frac{\partial }{\partial\phi}
\Bigr( \frac{H^{3/2}(\phi)}{8\pi^2}\, \frac{\partial \bigl({H^{3/2}(\phi)}
P_c(\phi,t|\phi_0) \bigr)}{\partial\phi}
\nonumber\\
 &+& \frac{V'(\phi)}{3H(\phi)} \, P_c(\phi,t|\phi_0)\Bigr)  \ .
\end{eqnarray}
For notational simplicity we took $M_p = 1$ in these equations.

One may try to find a stationary solution of equations (\ref{Starb}),
(\ref{E3711}), assuming that $\frac{\partial P_c(\phi,t|\phi_0)}{\partial t} =
0$. The simplest stationary solution (subexponential factors being omitted)
would be \cite{Star,Mijic,LLM}
\begin{equation}\label{E38a}
P_c(\phi,t|\phi_0) \sim   N~
\exp\left({3M_p^4\over 8 V(\phi)}\right)\cdot \exp\left(-{3M_p^4\over 8
V(\phi_0)}\right) \ .
\end{equation}
 The first term  in
this expression is equal to the square of the Hartle-Hawking wave function of
the universe (\ref{Vil1}), whereas the second one gives the square of the
tunneling wave function (\ref{E36}); $N$ is the overall normalization factor.
This result was obtained without any
ambiguous considerations based on the Euclidean approach to quantum cosmology.

This result has an obvious similarity with the Hawking-Moss expression for the
probability of tunneling, Eq. (\ref{HMinst}).   It provides a
simple interpretation of the Hawking-Moss tunneling. During inflation, long
wavelength perturbations of the scalar field freeze on top of each other and
form complicated configurations, which, however, look almost homogeneous on the
horizon scale $H^{-1}$. If originally the whole universe was in a state
$\phi_0$,
the scalar field starts wondering around, and
eventually it reaches the local maximum of the effective potential at $\phi =
\phi_1$. The probability of this event (and the typical time that it takes) is
suppressed by $\exp\left({3M_p^4\over 8 V(\phi_1)}\right)$. As soon as the
field $\phi$ reaches the top of the effective potential, it may fall down to
another minimum, because it looks nearly homogeneous on a scale of horizon, and
gradients of the field $\phi$ are not strong enough to pull it back to
$\phi_0$. This is  not a homogeneous
tunneling, but rather an inhomogeneous Brownian motion, which, however,  looks
homogeneous on the scale $H^{-1}$ \cite{book}. An important lesson is that
when one finds an instanton in de Sitter space describing homogeneous
tunneling, one should not jump to a conclusion that it really describes
creation of a homogeneous universe rather than an event which only looks
homogeneous on a scale $H^{-1}$.

\begin{figure}[Fig00]
 \hskip 1.5cm
\leavevmode\epsfysize=5cm \epsfbox{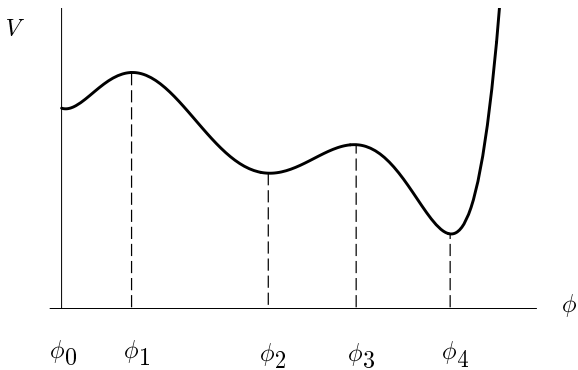}

\

\caption[Fig1]{\label{Fig00} Tunneling from the minimum at $\phi_0$ occurs not
to the points $\phi_2$ or $\phi_3$, which, according to the naive estimates
based on the instanton action, would be much more probable, but to the nearby
maximum at $\phi_1$.}

\end{figure}

That is how the stochastic approach resolves all mysteries associated with the
Hawking-Moss tunneling.  I believe that it is a very important point which
deserves a more detailed discussion. Consider for example the potential
$V(\phi)$ shown in Fig. \ref{Fig00}. There are five different de Sitter
instantons with action $S = -{3M_p^4\over 8 V(\phi)}$, corresponding to each of
the five extrema of this effective potential. How one should interpret them? Do
they describe tunneling between different minima, as suggested by Fig.
\ref{HawkMoss}, or creation from nothing, which can possible be described by
the upper half of Fig. \ref{HawkMoss}?

Thee are two ways of interpreting instantons. The first one is to say that they
{\it interpolate } between two different Lorentzian configurations, and
describe the tunneling between them. Then one should specify initial and final
states. This was the approach of Coleman and De-Luccia. The second one is to
avoid any discussion of tunneling (and creation) but simply use instantons as a
tool which allows to calculate the wave function of the ground state. This was
the approach of Hartle and Hawking.

As we already mentioned, using Hawking-Moss instantons as interpolating
Euclidean solutions is difficult (because each of these instantons describes a
constant field $\phi$), but   not impossible, see Fig. \ref{HawkMoss} and Ref.
\cite{Gonch}. Suppose we study tunneling from $\phi_0$ to $\phi_1$. Then the
probability of tunneling is given by Eq. (\ref{E38a}), where instead of $\phi$
one should use $\phi_1$. As one could expect, this result can be represented as
$e^{\rm \bf \Delta S_{01}}$, where ${\rm \bf \Delta S}$ is the change of
entropy between the initial and the initial states of the  system, ${\rm \bf
\Delta S_{01}} = {3M_p^4\over 8 V(\phi_1)} -{3M_p^4\over 8
V(\phi_0)} < 0$.

However, one could argue that it is much more probable to tunnel directly to
$\phi_2$, or to $\phi_3$, or to $\phi_4$. Indeed, the Hawking-Moss instantons
corresponding to each of these states do exist, and the absolute values of
their actions are much greater than of the action corresponding to the
tunneling to $\phi_1$. Naively, one would expect, for example, that the
probability of tunneling from $\phi_0$ to $\phi_3$ would be given by $e^{\rm
\bf \Delta S_{03}}$, where  ${\rm \bf \Delta S_{03}}= {3M_p^4\over 8 V(\phi_3)}
-{3M_p^4\over 8
V(\phi_0)} > 0$.  Of course, the probability of tunneling greater than $1$ does
not seem to make much sense, but this is what we get is we uncritically use the
Euclidean approach to tunneling. This is what one would expect in accordance
with the  argument of Ref. \cite{HTnew} implying that the universe should be
created in the state with the greatest entropy,  even if one encounters
suspicious expressions like $e^{\rm \bf \Delta S}$ with $ {\rm \bf \Delta S} <
0$.

{}From the point of view of the stochastic approach to inflation, the
resolution of the paradox is pretty obvious and quite instructive. First of
all, there is a subtle difference between the probability of tunneling and the
probability to find the universe in a state with a given field $\phi$. (This
issue is directly related to the difference between the instantons
interpolating between two different states, which describe tunneling,  and the
instantons used for the calculation of the wave function of the ground state.)
Strictly speaking, Eq. (\ref{E38a}) describes a stationary distribution of
probability to find a part of the universe in a state with a field $\phi$. It
does not necessarily describe  the  probability of tunneling (diffusion) to
this state $\phi$ from the state   $\phi_0$. These issues are  related to each
other, but only for $\phi \leq \phi_1$. According to \cite{Star,Star2,book},
the typical time which is necessary for the field  to move  from the local
minimum at  $\phi_0$   to any   field $\phi \leq \phi_1$  by the process of
diffusion is inversely proportional to $P_c(\phi)$.  That is why the
probability of jumping to the top of the barrier and roll down to $\phi_2$ is
proportional to $P_c(\phi_1)$. However, the probability distribution
$P_c(\phi)$, which is given by the square of the Hartle-Hawking wave function,
has no direct relation to the probability of tunneling to  $\phi > \phi_1$.
Once the field $\phi$ rolled over the barrier, the probability of its
subsequent rolling to $\phi_2$ is neither suppressed nor enhanced by any
additional factors.

Thus, despite expectations based on  the naive interpretation of the Euclidean
approach to tunneling, the universe does not jump to the state $\phi_2$, or
$\phi_3$, or $\phi_4$ with the probability greater than $1$. If initially the
main part of the universe  was in a state $\phi_0$, then the process of
diffusion gradually brings the scalar field $\phi$ in some parts of the
universe to the nearby maximum of the effective potential at $\phi_1$. The
probability of this event is {\it suppressed} by  $e^{\rm \bf \Delta S_{01}}
<1$. Then the field $\phi$ falls to the minimum at $\phi_2$. Diffusion from
$\phi_2$ to $\phi_3$ is also possible, and the probability to climb to $\phi_2$
is {\it suppressed} by  $e^{\rm \bf \Delta S_{23}} <1$. As a result, the
probability of diffusion (tunneling) from $\phi_0$ to $\phi_4$ is not enhanced
by $e^{\rm \bf \Delta S_{04}} > 1$ or by $e^{\rm \bf \Delta S_{03}} > 1$, as
one would naively expect, but is suppressed by
$e^{\rm \bf \Delta S_{01}}    e^{\rm \bf \Delta S_{23}} <1$.

Suppose now that we were waiting for a very long time, so that the scalar field
tunneled to its ground state $\phi_4$ in the main part of the comoving volume.
Then the probability distribution gradually reaches its stationary limit given
by Eq. (\ref{E38a})  provided that $V'' \ll H^2$ and inflation is possible all
the way from $\phi_0$ to $\phi_4$. This is exactly the result given by the
Hartle-Hawking wave function: The probability to be in a stationary state with
a small value of $V(\phi)$ is much greater than the probability to stay at
large $V(\phi)$. But, as we have seen,  this result has no direct relation
either to the probability of tunneling to the state near the absolute minimum
of the effective potential, or to creation of the universe in the state with a
smallest possible vacuum energy density.

One could argue that Eq. (\ref{E38a}) gives us much more than an interpretation
of
the Hawking-Moss tunneling. It appears to provide a direct confirmation and a
simple physical
interpretation of both the Hartle-Hawking wave function of the universe {\it
and} the tunneling wave function. First of all, we see that the distribution of
probability to find the universe in a state with the field $\phi$ is
proportional to $\exp\left({3M_p^4\over 8 V(\phi)}\right)$.  Note that we are
speaking here about the {\it state} of the universe rather than the probability
of its
{\it creation}.  Meanwhile, the probability that the universe emerged from the
state
with the field $\phi_0$ is proportional to $\exp\left(-{3M_p^4\over 8
V(\phi_0)}\right)$. Now we are speaking about the probability that a given part
of the universe was created from the state with the field $\phi_0$, and the
result coincides with our result for the probability of the quantum creation of
the universe, eq. (\ref{E36}).

This would be a great peaceful resolution of the conflict between the two wave
functions. Unfortunately, the situation is even more complicated. In all
realistic
cosmological theories, in which $V(\phi)=0$ at its minimum, the Hartle-Hawking
distribution $\exp\left({3M_p^4\over 8 V(\phi)}\right)$ is not normalizable.
The
source of this difficulty can be easily understood: any stationary distribution
may exist only due to the compensation of the classical flow of the field
$\phi$
downwards to the minimum of $V(\phi)$ by the diffusion motion upwards. However,
the
diffusion of the field $\phi$ discussed above exists only during inflation.
There is no diffusion
upwards from the region near the minimum of the
effective potential where inflation ends. Therefore the expression (\ref{E38a})
is
not a true solution of the equation (\ref{E3711}); all physically acceptable
solutions for $P_c$ are non-stationary (decaying) \cite{b19}.

One can find, however, stationary solutions describing the probability
distribution $P_p(\phi,t|\phi_0)$ introduced in \cite{b19}. This probability
distribution takes into account different speed of exponential growth of the
regions filled with different values of the field $\phi$. The investigation of
this
question shows \cite{LLM}, that the relative fraction of the volume occupied by
the field $\phi$ is described by a very complicated function which is
completely
different from the square of the Hartle-Hawking wave function. Meanwhile the
relative fraction of the volume of the universe from which any given part of
the
universe could originate is given by a function, which in the limit $V(\phi_0)
\ll M_p^4$ coincides with the square of the tunneling wave function. This is
additional evidence indicating that if one wants to find out where our part of
the universe came from, the investigation of the tunneling wave function can be
very
useful. As we already mentioned, the Hartle-Hawking wave function can be very useful too if one applies
it
for the investigation of perturbations near a classical de Sitter background.
One
can use it for the investigation of density perturbations in inflationary
universe
\cite{HHAL}, and for the study of black hole formation in de Sitter space
\cite{BH}. One
can also use it for the investigation of tunneling from a quasistationary state
in
de Sitter space \cite{HM}.  However, so far we did not find any evidence that
the Hartle-Hawking wave function describes the probability of quantum creation
of the universe.

\section{Quantum creation of open universes}

\subsection{\label{inst}Instantons describing creation of open universes from
nothing}

We discussed the difference between the two types of wave functions at such
length in order to put the results of Hawking and Turok into perspective.
They
have studied instanton solutions in the theories with inflationary potentials
such as $\phi^2$ or $\phi^4$. Since in these theories the field
$\phi$ moves, the corresponding instanton  somewhat differs from the standard
de
Sitter instantons. In particular, it contains a singularity at which the scalar
field $\phi$ becomes infinitely large. However, this field only logarithmically
grows near the singularity. At the same time the scale factor rapidly
decreases, and  the integral  $  -\pi^2\int d\tau a^3(\tau) V(\phi)$  giving
the  action $S$ converges. When  the tunneling occurs to the point $\phi$ where
inflation is possible, the main contribution to the action is
given by the nearly constant value of the field $\phi$. As a result, the
instanton action almost
exactly coincides with the usual de Sitter action $-{3M_p^4\over 16 V(\phi)}$
which we discussed before.\footnote{Expression for the action given in Eq. (8)
of Ref. \cite{HT} coincides with this expression, though it looks slightly
different because the authors used reduced Planck mass, which is smaller than
the usual one by a factor $(8\pi)^{-1/2}$.} This, in fact, was the assumption
which was made in \cite{Creation} in the investigation of the probability of
quantum creation of inflationary universe in the chaotic inflation scenario, so
now this assumption is verified.

An  important observation made in \cite{HT} was the possibility to make
analytical continuation of this instanton solution not only in the closed
universe direction, but in the open universe direction as well. This is very
interesting and nontrivial though not entirely new or unexpected because de Sitter
space is known to have an amazing property of being simultaneously closed, open
and flat. A similar analytical continuation was employed in the paper by Coleman and De Luccia \cite{CL}. The possibility of analytical continuation of the instanton
containing
a homogeneous scalar field to the open inflationary universe implies the
possibility of creation of an open inflationary universe containing a
homogeneous field $\phi$. This was the basis of  recent models  describing quantum creation of an open universe \cite{Gott,BGT,Open,LidOc,HOpen}. 

Quantum creation of an open universe from nothing may seem to be entirely
forbidden by the arguments contained in the previous section. Indeed, we are
talking about creation from nothing of a universe containing { \it infinite}
energy. However, this may not be a real problem here. Let us remember that in
the theory of the open universe creation by bubble formation
\cite{Gott,BGT,Open}, the universe inside the bubble looks finite from the
point of view of an external observer, but it grows infinitely large in time.
Its total energy grows because false vacuum gives its energy to the expanding
bubble wall. Thus, from the point of view of an inside observer, we have   an
instantaneous creation of an open universe with infinite total energy of
matter. However, from the outside, the same process looks like a continuous and
quite legitimate process of bubble growth and energy transfer from the
surrounding de Sitter space.

Similarly, an open universe created by tunneling in the model of  \cite{HT}
does not appear
alone, but as a part of a singular inflationary universe.
At the
moment when this complicated space-time emerges as a result of tunneling, the
total volume of the part occupied by an open universe is vanishingly small, and
it grows only gradually.
However,  just like in the growing bubble case
described above, one can make  a certain coordinate transformation after which
one may describe a part of the created space-time  as an infinite open
universe.

This is an a very interesting possibility which deserves further
investigation   \cite{tit}. Several  comments are in order here. First of all, the
instantons of this type describe tunneling not only to the part of the
potential where inflation is possible, but to non-inflationary parts as well.
In this case the scalar field may rapidly change with the growth of the
parameter $\tau$, and the action is not given by $-{3M_p^4\over 16
V(\phi(0))}$. Still the general tendency remains the same for all models we
analysed numerically: The smaller the potential
$V(\phi)$ at $\tau = 0$ (which corresponds
to the initial value of $V(\phi)$ in the open universe), the greater the
absolute value of the action.

\begin{figure}[Fig02]
 \centering
\hskip - 0.5cm
\leavevmode\epsfysize=14cm \epsfbox{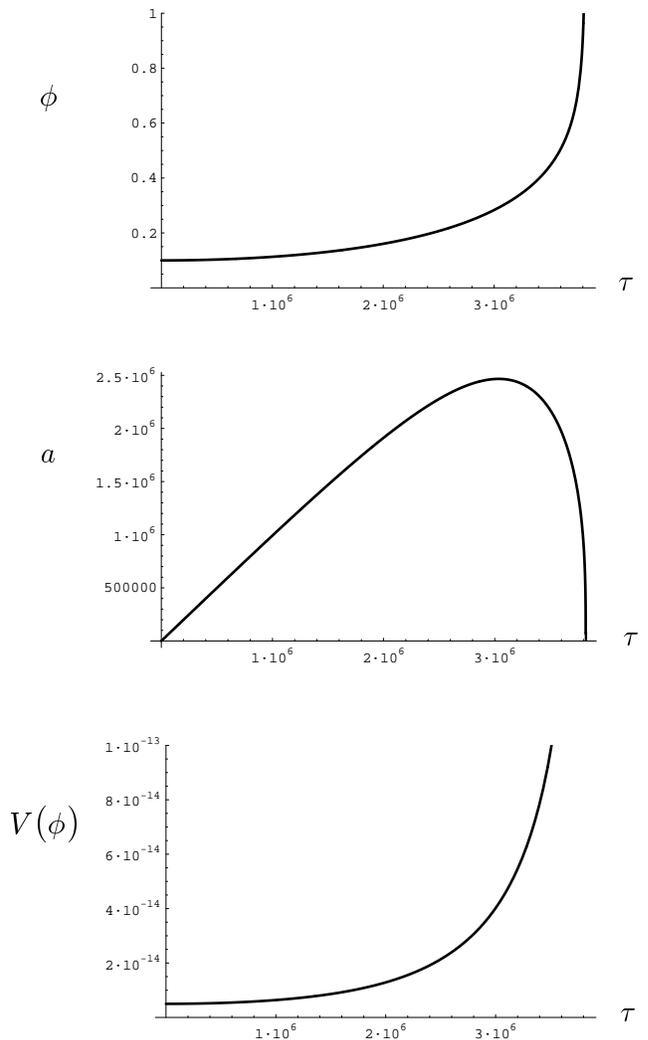}

\

\caption[Fig2]{\label{Fig02} Instanton in the theory ${m^2\over 2} \phi^2$
describing creation of the universe with $\phi = 0.1 M_p$. In this case the
scalar field rapidly changes in the Euclidean space, and the universe does not
inflate at all after the tunneling. If one considers greater values of $\phi$
at $\tau = 0$, the scalar field becomes almost constant, but then it diverges
logarithmically when   $\tau$ approaches its maximal value.}
\end{figure}

As an example, Fig. \ref{Fig02} shows the behavior of $\phi(\tau)$, $a(\tau)$
and
$V(\phi(\tau))$ for the instanton describing the universe creation in the
theory ${m^2\over 2} \phi^2$. We consider the case when $\phi(0) = 0.1$ in
units of $M_p$. In this case the universe does not inflate at all after the
tunneling. Still the instanton does exist. Its action is given by $S =
-\pi^2\int d\tau a^3(\tau) V(\phi)$. For the realistic value $m \sim 10^{-6}
M_p$ the  action for this case shown in Fig. \ref{Fig02}  is given by $-6.55
\times
10^{14}$, which is greater by an order of magnitude than the absolute value of
$-{3M_p^4\over 16 V(\phi(0))}$.
However, as soon as we consider tunneling to
the inflationary part of the effective potential, the function $\phi(\tau)$
becomes nearly flat (until it blows up near the singularity), and the action
practically coincides with the action on the usual de Sitter instanton with the
constant energy density $V(\phi(0))$: $S = -{3M_p^4\over 16 V(\phi(0))}$.

\begin{figure}[Fig03]
 \hskip 1.5cm
\leavevmode\epsfysize=14cm \epsfbox{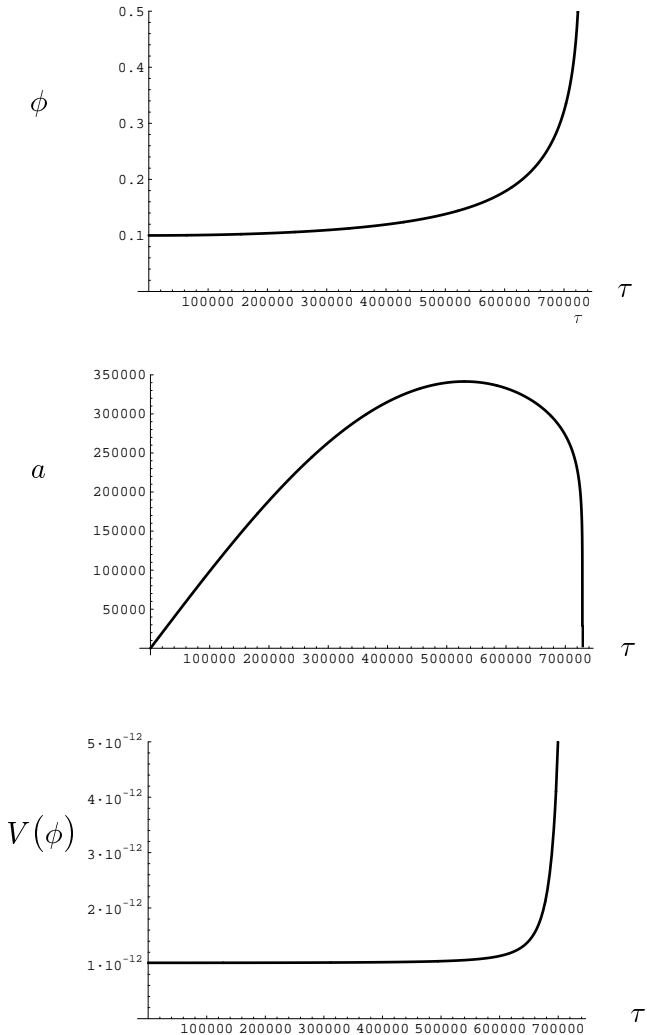}

\

\caption[Fig3]{\label{Fig03} Instanton in hybrid inflation model based on
supergravity, with $V(\phi) = M^4  (1 - Q\phi^2 + Q^2\phi^4)\, e^{Q\phi^2} $.
Everything is expressed in Planck units; $M \sim 10^{-3}$, $Q = 4\pi$, see Ref.
\cite{Riotto}.}
\end{figure}

In Fig. \ref{Fig03} we show the instanton in the theory with the effective
potential
$V(\phi) = M^4  (1 - Q\phi^2 + Q^2\phi^4)\, e^{Q\phi^2} $, with $M \sim
10^{-3}$, $Q = 4\pi$. This is the potential which (up to radiative corrections)
appears in the hybrid inflation scenario in supergravity proposed in
\cite{Riotto}, see also \cite{Papan}.   Note that the potential in this theory
is extremely steep at $\phi > 0.3$. Therefore inflation is possible only for
$\phi < 0.3$. Still the instanton solution does exist in this case as well.

In such theories, just like in the theories where the effective potential is
less steep, the field $\phi$ grows logarithmically near the singularity. If the
effective potential depends on the field exponentially, the contribution to the
action may blow up   there. The integral still converges (or diverges only
logarithmically) because of the sharp
decrease of $a(\tau)$ near the singularity.
It may still be necessary to make a cutoff of the integral, as soon as
the sharply growing function $V(\phi)$ becomes greater than $M_p^4$, and the
semiclassical approximation breaks down.

Thus, now we have a candidate for a new  mechanism of creation
of an open universe in
inflationary cosmology.
 There are many questions associated with the new instantons. First of all,
even though the singularity of the scalar field on these instantons is only
logarithmic, the singularity of the energy density and of curvature is
power-law. If one takes such instantons into account, the corresponding method
can no longer be called ``the no-boundary proposal.''  According to
\cite{VIL,BL}, the boundary terms give a contribution to the total action   $-{\pi M_p^2(a^3)'\over 4}$. We will not consider this correction  here;   it is
relatively small if one considers creation of an inflationary universe. For
example, the results of a numerical investigation performed in \cite{BL} show
that in the theory ${m^2\phi^2\over 2}$ this correction, as compared to the
action $-{3M_p^4\over 16 V(\phi)}$,  is suppressed by the factor $O({M_p\over
\phi})$ for $\phi \gg M_p$.

Another problem associated with the interpretation of the Hawking-Turok instanton as describing creation of an open universe
 was given recently in \cite{VIL}, and was related to the singular nature of the instanton. While we tend to agree with the main
conclusion of Ref. \cite{VIL}, we  do not think that every instanton having a singularity is disallowed.

 If one
considers the Hawking-Turok instantons and cuts them at the time of the maximal
expansion, $a = a_{max}$, they will look almost exactly like the Hawking-Moss
instantons. One may interpret them by saying that, just like in Fig. 1, they
interpolate between two realizations of a closed de Sitter space. If one
considers only the upper half of Fig. 1, starting from $a = 0$,   and making
the analytic continuation to de Sitter space at $a = a_{max}$, then one may
argue that the Hawking-Turok instanton interpolates between the state $a = 0$
(``nothing'') and de Sitter space. In such a case one may try to interpret this
instanton as describing creation of a closed universe from nothing. The results will not differ much if one calculate the action on a singular or on the nonsingular part of the instanton.

However, if one cuts the Hawking-Turok instanton  not by the plane $a =
a_{max}$, but by the plane  going through $a = 0$ and the singularity, as proposed in
Ref. \cite{HT} in order to describe creation of the open universe, it becomes much less obvious whether such an instanton
interpolates between any two well defined Lorentzian states, or even between
such  states as the state with $a = 0$ and the singularity. This
half-of-an-instanton seems to interpolate between half-of-nothing and
half-of-singularity. Thus we are not quite sure that it really describes
quantum creation of an open universe.  If the singularity is cut in half, and one needs to have a detailed knowledge of its structure to perform the analytical continuation, the possibility to use such instantons for the description of tunneling becomes very suspicious. For a more detailed discussion of this issue see \cite{BL}.

In addition, we have a general problem emphasized in the previous section. As we have found, only one of the instantons of  the Hawking-Moss type really describes  the tunneling  (the instanton describing the tunneling from $\phi_0$ to $\phi_1$), whereas all other instantons are irrelevant even though they are perfectly nonsingular. It is very hard to find any reason to discard them within the Euclidean approach to tunneling, but stochastic approach to inflation immediately explained which of them is relevant and what is its interpretation. We have found that it describes an inhomogeneous tunneling even though the instanton looks perfectly homogeneous.

We have a similar problem with respect to the Hawking-Turok instanton.   So far we were unable to find any interpretation of creation of a homogeneous open universe   within the stochastic approach to inflation. The closest thing we were able to find was the nonperturbative effect of creation of huge voids due to nonperturbative effects which might appear in a self-reproducing inflationary universe \cite{infloid}. However, this effect appears only if one introduces some specific probability measure in inflationary cosmology, related to the probability distribution $P_p$ mentioned in the previous section. Meanwhile the Hartle-Hawking wave function is related to the probability distribution $P_c$.

Despite all these problems, in what follows we will  make an assumption that the Hawking-Turok instantons do describe creation of a
homogeneous open universe, and that the corresponding action with a good
accuracy is given by $-{3M_p^4\over 16 V(\phi)}$. We will study consequences of
this assumption  if one interprets it using either the Hartle-Hawking or the
tunneling wave function. But one should remember that the validity of this
assumption made in \cite{HT} is less than obvious.

\subsection{\label{HHf}Open universes and the Hartle-Hawking wave function}

Possible implications of the new class of instantons depend crucially on the
choice of the wave function of the
universe.
Hawking and Turok suggested to use the Hartle-Hawking wave function, which
implies that the probability of the quantum creation of an open universe with a
field $\phi$ is given by Eq. (\ref{HH}):
\begin{equation}\label{HH2}
P \sim e^{-2S} = \exp \left({3 M_p^4\over 8 V(\phi)}\right) \ .
\end{equation}

This means that a typical open universe created by such a process would have
the
smallest
possible value of the field $\phi$, i.e. the universe would tend
to be created
directly in the absolute minimum of the effective potential, which does not
lead to any inflation whatsoever. As a result, such
a universe at present would be empty and would have  $\Omega = 0$. Of course
we cannot live in a universe with $\Omega = 0$, so
we should discard the universes with too small $\Omega$. Thus one may argue
that the final probability distribution to live in a universe with a given
value of $\Omega$ should be proportional to the product of the probability of
creation of such a universe and the probability of galaxy formation there. An
estimate of the most probable
value of $\Omega$ which one can observe with an account taken of anthropic
considerations can be made along the lines of \cite{E}. This estimate has lead
the authors of \cite{HT} to the conclusion that $\Omega$
should be about $10^{-2}$, which would be in a  disagreement with
the observational data suggesting that $\Omega \gtrsim 0.3$.

Is there a chance that this disagreement might disappear after a more detailed
investigation of the anthropic constraints on $\Omega$? After all, we are
talking
only about one order of magnitude, so  is it perhaps possible to make things
work? Let us consider this issue
more carefully.

One can parameterize the present value of $\Omega$ for an open universe in the
following way \cite{HT}:
\begin{equation}\label{OMEGA}
\Omega \approx {1\over 1+ A e^{-2N(\phi)}} \ ,
\end{equation}
where $A$ is some factor depending on the efficiency of reheating and other
details of the theory, and $N(\phi)$ is the number of e-folding of inflation
after the field $\phi$ begins rolling down.

Let us compare the probability $P_{\phi}$ for the universe to begin at some
value of the scalar field $\phi$, and the corresponding probability to have a
slightly greater field $\phi + \Delta \phi$:
\begin{equation}\label{PROB}
{P(\phi)\over P(\phi+\Delta\phi)} = \exp \left({3 M_p^4 V'\Delta\phi \over 8
V^2(\phi)}\right).
\end{equation}
Now one should take into account that $\Delta N = H
\Delta t = {H \Delta \phi\over \dot \phi} = {3H^2 \Delta
\phi\over V'}$, i.e. $\Delta\phi = {\Delta N V'
M_p^2\over 8\pi V }$. Also, the amplitude of density perturbations $\delta \sim
{M_p^3V' \over V^{3/2}} \sim 10^{-5}$. Combining this all together and dropping
factors $O(1)$, one has
\begin{equation}\label{PROB2}
{P(\phi)\over P(\phi+\Delta\phi)} = \exp
\left(10^{-1}\delta^{-2}\Delta N \right) \sim \exp
\left(10^{9} \Delta N \right).
\end{equation}

The universe with $\Omega = 0.3$ appears after the creation of the universe
with
$\phi_{0.3}$, where $0.3 = {1/ (1+ A e^{-2N(\phi_{0.3})})}$. This gives $A
e^{-2N(\phi_{0.3})}
\approx 2$. Meanwhile the universe with $\Omega = 0.2$
appears if $A e^{-2N(\phi_{0.2})} \approx 4$. Therefore $\Delta N =
N(\phi_{0.3}) - N(\phi_{0.2}) \sim 0.5$. Thus the probability of creation of a
universe with $\Omega = 0.2$ is approximately $10^{10^8}$ times greater than
the
probability of creation of the universe with $\Omega = 0.3$. Clearly, the
probability of galaxy formation in these two cases cannot
differ by a factor $10^{10^8}$. This means that according to \cite{HT} it is
entirely improbable to live in the universe with $\Omega = 0.3$. Similarly, it
seems entirely improbable to live in the universe with $\Omega = 0.2$. One
should consider absolutely extreme conditions in the universe in order to
compensate for the factors of the type of $10^{10^8}$. Note that this
conclusion
is valid independently of the choice of the inflationary potential: The final
result is determined by the amplitude of density perturbations $\delta$ which
is
given by observations.

Since the probability of the universe formation grows roughly $10^{10^8}$ times
when the number of e-foldings $N$ decreases by $\Delta N = O(1)$, this growth
can be compensated only by decreasing the probability of galaxy formation at
small $N$ (at small $\Omega$). To compensate the factor $\sim 10^{10^8}$, the
probability of galaxy formation must be smaller than $10^{-10^8}$. In such
a
case we would not see any other galaxies around us; the next nearby galaxy in
the Hartle-Hawking universe would be at a distance about $10^{100000000}$
light
years away...

One could hope that in the worst case one can simply  return to the old method
of creation of the open universe
proposed in \cite{Gott,BGT,Open}. However, once
the Hartle-Hawking approach is adopted, this does not seem possible either.
The main difference between the previous mechanism of the open universe
formation and the new one was the existence of a deep local minimum of the
effective potential at some $\phi = \tilde\phi$. In such a situation there
exists the Coleman-De-Luccia instanton, which describes the creation of an open
universe immersed in the false vacuum with $\phi = \tilde \phi$. The minimal
value of the scalar field $\phi$ on this instanton solution $\phi_{\rm min}$
should still be sufficiently large for the further 60 or 70 e-foldings of
inflation to occur inside the bubble. This, if we would try to obtain an
instanton solution for such theories just like we did for several other
theories, see Figs. 1 and 2, we would begin our calculations at $\phi(0)
=\phi_{\rm min}$, and we would see that the growing   field $\phi$ stabilizes
at $\phi  = \tilde \phi$. However, if one starts the calculations at $\phi <
\phi_{\rm min}$, the scalar field   rolls over the local minimum of the
effective potential (which looks like a local maximum from the point of view of
equations of motion in Euclidean space) and  continues its growth toward
indefinitely large $\phi$.  Thus the Hawking-Turok instantons do exist even in
the theories where the effective potentials grow nonmonotonically at large
$\phi$. Therefore the conclusion concerning the tunneling to smallest possible
values of $V(\phi)$ in the context of the Hartle-Hawking approach seems to be
quite general.

In the theories with two fields of the type of
(\ref{4a}) the situation is even easier to analyse: One may consider the
instanton with $\phi = 0$ describing the field $\sigma$ climbing from the
minimum of the effective potential to $\sigma \to \infty$. All results obtained
in \cite{HT} and above apply to this case.

Thus, we expect that the instantons of the type considered above should exist
in the models considered in  \cite{Gott,BGT,Open}, and therefore   all
conclusions about the preferable creation of the universe with the smallest
possible $V(\phi)$ and extremely small $\Omega$ should apply to such theories
as well. This implies that the Hartle-Hawking approach makes it extremely
difficult to propose {  any} realistic open inflation model.

What happens if one makes a different analytical continuation and concentrates
on the closed universe case instead? Similarly, the probability of quantum
creation of the universe grows $10^{10^8}$ times if the duration of inflation
becomes one e-folding shorter. But a closed universe which inflates less than
$N
\sim 70$ (the exact number depends on the features of reheating) collapses in
less
than $10^{10}$ years, which makes the existence of life very problematic.
Again,
the only way to compensate for the factors $\sim 10^{10^8}$ pushing the
probability distribution toward small $N$ (i.e. toward the premature death of
the universe) is to assume that the probability of the existence of life near
an
``optimal'' $N$, corresponding to the maximum of the total probability
distribution, decreases by more than $10^{-10^8}$ when $N$ decreases by $1$.
Thus, the no-boundary proposal based on the Hartle-Hawking wave function
pushes us toward  the region where the existence of life becomes nearly
impossible. This does not mean that this
proposal is incorrect. As we have
already argued, it works
perfectly well if one calculates the probability of events produced by usual
quantum mechanical fluctuations having positive energy near an already existing
cosmological background. However, we believe that it does not apply for the
calculation of the probability of formation of this background, which involves
the investigation of the
fluctuations of the scale factor. If one does not make an attempt to extend the
validity of the Hartle-Hawking wave function beyond a certain point, one does
not face the consequences discussed above.

One possibility to resolve this problem is suggested by the form of the boundary terms found in   \cite{VIL,BL,HT2}. For example, in chaotic inflation with $V(\phi) = {m^2\over 2}\phi^2$ the action with an account taken of boundary terms is given  by \cite{BL}:
\begin{equation}
S 
\approx
-{3 M_{\rm p}^4\over 8 V(\phi)}\left(1 - {M_{\rm p} \over
2\phi}\right).
\end{equation}
This equation shows that the   action becomes minimal not at $\phi = 0$, but at   $\phi\sim M_p$, which allows for a short stage of inflation. However, a numerical investigation of this question performed in \cite{BL} for several different versions of chaotic inflation scenario has shown that this stage of inflation is extremely short, and the corresponding value of $\Omega$ would be exponentially small.

Hawking and Turok proposed to consider an inflationary model with a local maximum of the effective potential, such as  $V(\phi) =\mu^4(1-{\rm cos}(\phi/v))$ \cite{HT2}. In this model the top of the effective potential corresponds to a local minimum of the action with an account taken of boundary terms. If one neglects   the possibility of tunnelling to small $\phi$, the second best possibility is that the universe is created at the top of $V(\phi)$.
But if the total duration of inflation is small (which is necessary to keep the universe open), then the tunneling to the top is not allowed. Indeed,   density perturbations produced during inflation are inversely proportional to $V'$, so they are very large at the top of $V(\phi)$.   Large amplitude of density perturbations on the horizon  is anthropically forbidden, so it was argued in \cite{HT2} that we should tunnel not to the top but to some point $\phi^*$ away from the top. This hopefully could give  $\Omega \sim  0.3$ and ${\delta\rho \over \rho} \sim 10^{-5}$ on the horizon, after a certain fine-tuning of the parameters of the model.

But suppose that we indeed have a model with parameters which make it possible. Then in the same model it is even more probable to tunnel directly to the top (because the action is smaller there), and then roll to $\phi^*$ from the top. We will still have ${\delta\rho \over \rho} \sim 10^{-5}$ on the horizon, but in this case we will have $\Omega = 1$ because of the additional stage of inflation during the rolling from $\phi = 0$ to $\phi = \phi^*$.  This is not anthropically forbidden because ${\delta\rho \over \rho}$ only very weakly depends on the length scale; it becomes much greater than $10^{-5}$  only at distances much greater than the size of the observable part of the universe. Thus it does not seem possible to get $\Omega < 1$ in this version of the scenario proposed in Ref. \cite{HT2} even if one uses the Hartle-Hawking wave function and take into account boundary terms.

Until now we tacitly assumed that the creation of the universe is a one-time event, and that it is  correct to describe the total probability of forming
a galaxy
as a product of the probability of creating the universe with a given
$\phi$ and the probability of forming a galaxy for a given $\Omega(\phi)$.
This is a reasonable proposal in the minisuperspace approach to quantum
cosmology, but it may fail if one takes into account the effect of
self-reproduction of the universe. Indeed, the probability of creation of the
universe with a large field $\phi$ is very small in the context of the
Hartle-Hawking proposal. However, the universes with large $\phi$ in the
chaotic inflation scenario typically enter the stage of eternal
self-reproduction, which leads to a permanent exponentially rapid growth of their total volume \cite{b19}. This process  leads to creation of infinitely large number of galaxies. Then a typical galaxy will be produced not in the region suggested by the
Hartle-Hawking probability distribution, but in the region where the scalar
field $\phi$ was large enough for the process of self-reproduction of the
universe to begin. 

 One could object that if the Hartle-Hawking wave function correctly describes creation of an open universe,  then the universe has very small energy density from the very beginning, and self-reproduction of the universe never happens. However, according to \cite{LLM}, if the universe is  sufficiently large, the process of self-reproduction occurs   even if the initial value of the field $\phi$ is so small that it can barely support inflation. Thus self-reproduction definitely occurs inside an infinite open inflationary universe. In such a case all negative (and positive)
consequences of the description of quantum creation of the universe by the
Hartle-Hawking wave function disappear, not because the consequences of the
no-boundary proposal become different, but because the choice of initial
conditions in quantum cosmology provided by the Hartle-Hawking (or tunneling)
wave function   becomes irrelevant for the description of the
main part of our universe \cite{Mijic,LLM,Page}. In the first universe produced by quantum creation  from nothing one may have $\Omega \sim 0.01$, if it is described by the Hartle-Hawking wave function, or $\Omega = 1$, if it is described by the tunneling wave function. However, this universe will produce infinite number of new inflationary universes. One may wonder, what is the most probable origin of a  part of the universe of a given physical volume, which has density $\rho$ at the time $t$  after the creation of the universe from nothing?
The answer is that  the relative fraction of the physical volume of a self-reproducing universe
in a state  with given properties (with  given values of  fields,
with a given density of matter, etc.)  does not depend on time $t$. The probability   that a given part of the universe in this scenario originated from a state with a certain value of the scalar field $\phi$ is given by a function which is very similar to the square of the tunneling wave function \cite{LLM}.

\subsection{\label{tunn}Open universes and the tunneling wave function}

Now let us see what happens if we use the results of Ref. \cite{HT}, but
interpret them from the point of view of the tunneling wave function. In this
case, according to \cite{Creation,ZelStar,Rubakov,Vilenkin}, the probability of
the universe
creation is proportional to $\exp \left(-{3 M_p^4\over 8 V}\right)$. Thus the
universe tends to be born at the highest possible value of the
effective potential $V$.
In the simplest models with
the effective potentials ${m^2\over 2}
\phi^2$ or ${\lambda\over 4} \phi^4$ the total duration
of inflation is so large that the resulting value of $\Omega$ becomes equal to
$1$ independently of the way the universe was born (i.e. whether it was closed
or open from the very beginning). One may or may not like it, depending on
one's
beliefs concerning the total density of the universe at present, but at least
this value is not as far away from the recent observational results as the
conclusion that we should live in a structureless universe with $\Omega =
0.01$.

On the other hand, now we have two classes of models where one can get $\Omega
<1$. The first class includes all models proposed in
\cite{Gott,BGT,Open,LidOc,HOpen}. The universe may begin in a singularity, or
it
may appear due to creation from nothing. The final result will be entirely
insensitive to it. Indeed, as soon as inflation begins, in most
versions of the
of chaotic inflation scenario the universe enters the regime of eternal
self-reproduction \cite{book}. It
produces an indefinitely large amount of space. For example, in the simplest
model with the potential (\ref{4a}), the eternal inflation may begin at
very large $\sigma$ and $\phi$ \cite{Open}. Then it produces exponentially
large domains
filled with all possible values of $\sigma$ and $\phi$. In particular, there
will be domains trapped in the local minimum near $\sigma = 0$. These domains
will continue to inflate eternally, like the de Sitter phase in the old
inflation
scenario, and they will continue producing open inflationary universes with all
possible values of $\Omega$. Thus, in this scenario a single act of creation of
the universe produces not one but an infinite number of open universes.

One may wonder what is the most probable value of $\Omega$ in this scenario. At
the moment we do not know a definite answer to this question. In all versions
of
eternal inflation theory we have to compare an infinite number of universes
with
different properties. As a result, the answer is ambiguous; it depends on the
way one performs a cut-off and regularizes the infinities. For a discussion of
different approaches to this question one may see, e.g., \cite{Open,LLM,VW,LM};
the problem is not settled yet. We do not even know whether it makes any sense
to look for a definite answer. The reason is very simple \cite{LLM,LM}.
Consider
two {\it infinite} boxes, one with apples, another with oranges. One can pick
one fruit from each box, an apple and an orange,
then again an apple and an orange, and so on.
This may
give an idea that the number of apples is equal to the number of oranges. But
one can equally well take each time one apple and two oranges, and conclude
that
the number of oranges is twice as large as the number of apples. The main
problem here is that we are making an attempt to compare two infinities, and
this gives an ambiguous result. Similarly, the total volume of a
self-reproducing inflationary universe diverges in the future. When we make
slices of the universe by hypersurfaces of constant time $t$, we are choosing
one particular way of sorting out this infinite volume. If one makes the
slicing
in a different way, the results will be different. The main statement, which
does not depend on the choice of the probability measure, is that we have an
infinite number of apples and oranges, and we have an infinite number of
domains
with various values of $\Omega$. If we want to find in which of these universes
we live, we should go and measure the value of $\Omega$; whichever we find will
be ours.\footnote{Inflationary theories of this type which allow a definite
prediction for $\Omega$ are also possible \cite{LidOc}.}

In addition to this class of theories, we may consider another class, which was
introduced in \cite{Open} when we studied the possibility of creation of a
closed universe with $\Omega$ substantially greater than $1$, see Introduction.
The main idea is to consider the models where self-reproduction of the universe
is impossible and the total duration of inflation is very small. For example,
one can consider a particular version of the chaotic inflation scenario with
the
effective potential
\begin{equation}\label{1}
V(\phi) = {m^2 \phi^2\over 2}\, \exp{\Bigl({\phi\over CM_p}\Bigr)^2} \ .
\end{equation}
Potentials of a similar type often appear in supergravity. In this theory
inflation occurs only in the interval ${M_p\over 2} \lesssim \phi \lesssim
CM_p$. One may consider a model of tilted hybrid inflation    proposed in
\cite{GBL}, or a particular version of hybrid inflation in supergravity
proposed in \cite{Riotto,Papan}, where the effective potential at large $\phi$
(when
logarithmic terms appearing due to quantum corrections become subdominant)
looks as follows:
\begin{equation}\label{RIOTTO}
V(\phi) = M^4  \left(1 - Q{\phi^2\over M_p^2} + Q^2{\phi^4\over M_p^4}\right)\,
\exp\left({Q\phi^2\over M_p^2}\right) \ .
\end{equation}
Here  $M \sim 10^{-3} M_p$, $Q = 4\pi$.   As we have
already mentioned in Sect. \ref{inst},  the effective potential in this theory
is extremely steep at $\phi > 0.3$. Therefore inflation is possible only for
$\phi < 0.3$. Still the
instanton solution does exist, both for $\phi(0) < 0.3$ and for $\phi(0) >
0.3$. All coupling constants in this model are
$O(10^{-1})$, and the total duration of inflation is $N
\sim 10^2$. This makes it an interesting candidate for the
open inflation model.

One may also consider models with the simple quadratic effective potential
${m^2\over 2} \phi^2$, but assume that the field $\phi$ has a
nonminimal interaction with gravity of the form $-{\xi\over 2} R\phi^2$. In
this
case inflation becomes impossible for $\phi > {M_p\over \sqrt{8\pi\xi}}$
\cite{Maeda,BL95}. In order to ensure that only a limited amount of inflation
is possible for
inflationary
universes which
can be produced during the process of quantum creation of the
universe
in the theory ${m^2\over 2} \phi^2$, it is enough to assume that ${M_p\over
\sqrt{8\pi\xi}} < 3M_p$.
This gives the condition $\xi > {1\over 72\pi} \sim
4\times10^{-4}$.

If an open universe is created and it does not inflate much, then after
inflation we have an open universe with $\Omega < 1$ in either of the models
described above.

There are several different problems associated with this scenario. Consider
for definiteness the model (\ref{1}) and suppose for a moment that the
tunneling may occur only to the region of small $\phi$, where inflation is
possible. Then, according to Eq. (\ref{TUNN}),
the maximum of probability of creation of an inflationary universe appears near
the upper range of values of the field $\phi$ for which inflation is possible,
i.e. at $\phi_0 \sim C M_p$. The probability of such an event will be so
strongly suppressed that the universe will be formed almost ideally homogeneous
and spherically symmetric. As pointed out in \cite{Lab}, this solves the
homogeneity, isotropy and horizon problems even before inflation really takes
over. Then the size of the newly born universe in this model expands by the
factor $\exp({2\pi
\phi_0^2M_p^{-2}})\sim \exp({2\pi C^2})$ during the
stage of inflation \cite{book}. If $C \gtrsim 3$, i.e.
if $\phi_0 \gtrsim 3M_p
\sim 3.6\times 10^{19}$ GeV, the
universe expands more than $e^{60}$ times, and it becomes very flat. Meanwhile,
for $C \ll 3$ the universe always remains ``underinflated'' and very curved.
Its
properties will depend on the way it was formed. If we make analytical
continuation of the Hawking-Turok instanton in the usual way, it will describe
a
formation of a closed universe with $\Omega > 1$. On the other hand, the new
analytical continuation proposed in \cite{HT} describes creation of an open
universe with $\Omega < 1$. In order to obtain $\Omega$ in the interval
between $0.3$ and $0.2$ at the present time one should have the constant $C$ to
be fixed somewhere near $C = 3$ with an accuracy of few percent. This is a
fine-tuning, but not a terrible one.

However, in the above analysis we have
assumed that the tunneling may occur only to
the inflationary part of the effective potential. Meanwhile we obtained
instanton solutions which describe tunneling to noninflationary parts of the
effective potential as well, with $\phi_0 \gg C M_p$. This may not look
problematic. The field tunnels to the highest possible position with $V(\phi)
\sim M_p^4 $. Then after many oscillations the scalar field decreases to the
region where inflation becomes possible, then the universe inflates a little,
and we still get the universe with $\Omega < 1$. The problem is that if, as we
expect,  the  probability to create a universe with a nearly Planckian density
is not strongly suppressed, then at the moment of its creation the universe
will not be very homogeneous. If the universe inflates a lot after its
creation, these primordial inhomogeneities do not make us any harm. However, if
inflation produces the universe with $\Omega < 1$, these inhomogeneities may
cause significant anisotropy of the CMB radiation.

It may happen that this is not a real problem. A typical scale factor of an
open universe at the moment of its creation in this scenario will be
$O(M_p^{-1})$. Thus one may expect initial inhomogeneities to exist on this
scale. Then the  scale factor of the universe, as well as the wavelength of
these perturbations,     expands more slowly than the size of horizon $\sim t$
until the universe becomes inflationary. As a result, all  initial
inhomogeneities at the beginning of inflation have wavelengths much shorter
than the horizon. Such perturbations rapidly decrease during inflation and
become harmless. This may solve the homogeneity problem, but we believe that
this issue requires a more detailed investigation.

Note also that this problem appears only if we assume that the tunneling to
large values of   $V(\phi)$ is possible. But what if the scalar field $\phi$ is
only an effective degree of freedom describing, for example, the radius of
compactification, or a condensate of fermions?  Then the effective potential
may not be defined at $V(\phi) \sim M_p^4$, the tunneling to very large
$V(\phi)$ becomes impossible, and the homogeneity problem may disappear.

One more thing which should be analysed is the applicability of the simple rule
$P \sim e^{-|S|}$ for the description of the universe creation in the models
with steep potentials. Indeed, as we emphasized, we expect this expression to
be valid in the situations when one can neglect motion of the scalar field. In
this case one can treat $V(\phi)$ as a cosmological constant and quantize only
the scale factor. One may expect that this rule will remain approximately
correct if the motion of the field $\phi$ is very slow. But if the effective
potential is very steep, the field $\phi$ will move very fast. In such cases  one should quantize simultaneously the scale factor $a$, which has negative energy, and the scalar field $\phi$, which has positive energy. In this case the
relation $P \sim e^{-|S|}$ must be considerably modified. One such example is
the pre-big-bang cosmology, where action vanishes identically on equations of
motion, whereas the entropy of inflationary universe is exponentially large
\cite{KLB}.  In our case there is an additional modification related to the boundary terms, which become very significant for tunneling to the steep parts of the effective potential. Indeed, the numerical investigation of this issue performed in \cite{BL} shows that in the regions where the effective potential is very steep the boundary terms may become so large that they may even change the sign of the action. This simply implies that the naive expression for the tunneling wave function obtained by  modifying the sign of the action does not apply to such situations. However, this does not change our general qualitative conclusion
that tunneling with creation of the universe with $V(\phi) \sim M_p^4$ is not
suppressed.

In addition to all problems mentioned above, one should also  make sure
that the leading channel of the universe creation will produce
topologically trivial open universes. First of all, the
tunneling may produce closed
universes as well, with a similar probability \cite{Open}. This is not   a real
problem though, because if the tunneling occurs to small $\phi$, so that in
the open universe case one obtains the universe with $\Omega \ll 1$, then in
the
closed universe case the same instanton will describe the universe with a very
large $\Omega$ which collapses too early for any observers to appear there.
But creation of a closed universe is not the only competing process.
There exist a variety of instantons describing Euclidean universes
with a nonvanishing vacuum energy density $V$. The usual de Sitter instanton
discussed above is just one of them. For example, the action on the Page
instanton $P_2 + \bar P_2$ is $-{9 M_p^4\over 40 V}$, the action on the
Fubini-Study instanton $P_2$ is $-{9 M_p^4\over 32 V}$, the action on the $S^2
\times S^2$ instanton is $-{M_p^4\over 4 V}$ \cite{Eguchi}.\footnote{The values
of the action we gave here are obtained by integration over the full Euclidean
space, so they should be compared to the complete de Sitter action
$-{3M_p^4\over 8 V}$.}

The most interesting of these solutions is the $S^2
\times S^2$ instanton. The absolute value of its action
is smaller than that of de Sitter instanton, so one may argue that it is easier
to create an anisotropic Kantowski-Sachs universe rather than the isotropic de
Sitter space. Note that the resulting geometry is unstable with
respect to the exponential growth of the radii of both spheres, and
eventually this solution becomes locally indistinguishable from de Sitter space
\cite{SK}. However, if the tunneling occurs to small $\phi$, the universe does
not expand long enough to erase the large-scale anisotropy, which should
therefore
be detectable.

One should note, that it is not quite correct to directly compare the action of
de Sitter instanton to the action of the $S^2 \times S^2$ instanton. Indeed, in
the theories
where the effective potential sharply rises at large $\phi$, the action
describing the tunneling to large $\phi$ is not given by the simple expressions
of the type of $-{3 M_p^4\over 16 V}$, but should be calculated anew for each
particular configuration. If the tunneling occurs near the Planck density, its
probability is not expected to be exponentially suppressed for either of these
instantons, so the probability of creation of different spaces may be
comparable, and then we may live in the universe with the simplest topological
properties (if it is true) merely by chance.

It is also quite possible that the tunneling may create spaces of a more
complicated topology. The first attempt to study this possibility was made in
\cite{ZelStar}. It was found that the probability of tunneling to a flat
exponentially expanding space with identified sides may not be suppressed at
all unless one takes
into account quantum corrections to the energy momentum tensor. This space has
metric of a 3-torus with identified sides,
\begin{equation}\label{star}
ds^2 = dt^2 - (a^2(t) dx^2 + b^2(t) dy^2 + c^2(t) dz^2) \ ,
\end{equation}
with $x +L \equiv x$, $y +L \equiv y$, $z +L \equiv z$. At large $t$ this space
locally looks like de Sitter universe, but if the expansion is not long enough,
then
the universe will be noticeably anisotropic.

All these problems would not even arise in the standard situation when
inflation
lasts much more than $60$ e-foldings, but if one adjusts the parameters of the
model in such a way as to have inflation very short, the issue of global
anisotropy and topology of the universe becomes quite important, see in this
respect \cite{Cornish}.

Another potential drawback of the new class of open inflation models is the
unusual shape of the spectrum of density perturbations. By construction,
inflation in these models begins at the point when the slope of the effective
potential for the first time becomes not very steep, and the friction produced
by the term $3H\dot\phi$ for the first time becomes sufficient to slow down the
rolling of the
field $\phi$. But this automatically means that the amplitude of density
perturbations produced at the beginning of inflation, which now corresponds to
the scale of    horizon,
should be very small (blue spectrum), see e.g. \cite{Riotto}. This may be a
real
problem for such models. Note, however, that this problem is somewhat opposite
to the previously discussed problem of overproducing large scale density
perturbations created during the tunneling.

All these questions require a thorough investigation to make  sure that
the new    models of open inflation discussed above can work. As we already emphasized in Sect. \ref{inst}, we are not sure that the Hawking-Turok instanton really describes quantum creation of an open universe. 
It is important,
however, that quite independently of these new possibilities, which may or may
not prove to be realistic,   the tunneling wave function allows us to have
usual inflationary models predicting $\Omega = 1$,
as well as the previously proposed class of models with $\Omega < 1$
\cite{Gott,BGT,Open,LidOc,HOpen}. It seems much better than to have models
predicting either $\Omega \gg 1$ for the closed universe case, or $\Omega \sim
10^{-2}$ for the open inflationary universe.

\section{\label{F}Models with the antisymmetric tensor field}
In order to avoid the unfortunate consequence $\Omega \sim 10^{-2}$ of their original model, Hawking and Turok introduced recently a new class of models \cite{HT2}, where they added the four form field strength
$F_{\mu \nu \rho \lambda}= \partial_{[\mu}A_{\nu \rho \lambda]}$.  The Euclidean action for their model   is:
\begin{eqnarray}
{\cal S}_E  &=&  \int d^4 x \sqrt{g}\, \Bigl( -{1\over 16 \pi G} R +{1\over 2}
g^{\mu \nu}\partial_\mu \phi \partial_\nu \phi + V(\phi) \nonumber\\
 &+&{1\over 48}
F_{\mu \nu \rho \lambda} F^{\mu \nu \rho \lambda}\Bigr)  
 + {1\over 8 \pi G} 
\int d^3 x \sqrt{h} K \ ,
\label{newaction}
\end{eqnarray}
The last term gives the boundary contribution, which is typically small when the tunneling occurs to the values of $\phi$ corresponding to a long stage of inflation \cite{BL,HT2}.

The field $F$ in four-dimensional space is not a real dynamical field. The Lagrange equation  for $F$ in the Euclidean regime has  a solution $F^{\mu \nu \rho \lambda}= {c\over \sqrt{g}} \epsilon^{\mu \nu \rho \lambda}$ with $c$ an arbitrary constant. In the Lorentzian  regime this solution becomes $F^{\mu \nu \rho \lambda}= {ic\over \sqrt{-g}} \epsilon^{\mu \nu \rho \lambda}$. Its main role is to give a contribution to the effective cosmological constant, $V(\phi) \to V(\phi) - {1\over 48} F^2$, where $F^2 \equiv  F^{\alpha \beta \gamma \delta}  F_{\alpha \beta \gamma \delta}$. The trick is to add simultaneously the vacuum energy $V_0 =  {1\over 48} F^2$. This operation leaves the original value of $V(\phi)$ intact, and thus it  does not lead to any  effects if one calculates the entropy of the nearly de Sitter space, 
\begin{equation}\label{entropy}
{\rm \bf S} =
{3M_p^4\over 8 V(\phi)} .
\end{equation}   
Here by $V(\phi)$ we mean the total  energy density, including the energy of the scalar field $V(\phi) +V_0$ and the compensating $F^2$ contribution. For example, one can take  $V(\phi) = {m^2\phi^2\over 2} + V_0 -{1\over 48} F^2 = {m^2\phi^2\over 2} $.  

However, if instead one calculates the Euclidean action, which normally coincides with  $-{\rm \bf S}$ in inflationary cosmology, one gets a different result \cite{Duff}. The action becomes a nontrivial function of $V(\phi)$ and $F^2$. Neglecting the small boundary term, and integrating over the entire solution (which doubles the result), one gets \cite{Duff,HT2}:
\begin{equation}\label{modaction}
S \approx - {3M_p^4 \over 8 V(\phi)^2} (V(\phi)-{1\over 24} F^2 ) \ .
\end{equation}
This coincides with (minus) entropy $-{\rm \bf S}$ for $F = 0$. However,   for  $V_0 =  {1\over 48} F^2 \not = 0$ one no longer has the maximum of absolute value of action $S$ at $V(\phi) = 0$. Instead of that, the maximum is reached at $V(\phi) \sim 4 V_0$. By a proper choice of $V_0$ one can fine tune the most probable initial value of $\phi$ (according to the Hartle-Hawking prescription) to be at any given place. In particular, one can have it at $\phi \sim 3M_p$, which would lead to about 60 e-folds of inflation. Thus, by choosing the proper value of the constant $V_0$ one can obtain any value of $\Omega$,  from $0$ to $1$.

A few comments are in order here. The main reason why the original idea of Hawking and Turok was so attractive is the postulated absence of any fine-tuning. Now this is no longer the case. Consider for example a realistic model of chaotic inflation with  $V(\phi) =   {m^2\phi^2\over 2} $, with $m \sim 10^{-6} M_p$. To obtain the most probable value of $\phi$ near $3M_p$ one would need to have $V_0 \sim 10^{-12}$ in units of the Planckian energy density $M_p^4$. This introduces a new extremely small parameter to the theory. The value of this parameter $ {V_0\over M_p^4} \sim 10^{-12}$ must be further fine-tuned with an accuracy of about 1\% in order to get the desirable value of $\Omega$. Then $F^2$ should be fine-tuned to cancel $V_0$ with an accuracy  $10^{-123}M_p^4$, which is achieved in \cite{HT2} by using anthropic considerations.

This mechanism can work only if $F$ is imaginary in the Lorentzian regime. It is not quite clear therefore whether this model is realistic. 

An additional complication appears if one remembers that now the entropy no longer coincides with the (minus) Euclidean action. Thus, one may wonder which of these functions should be maximized? The extremum of entropy ${\rm \bf S}$, as before, appears at the point corresponding to the absolute minimum of  $V(\phi)$, independently of the presence of the field $F$. Thus, the argument that one should maximize the entropy, given in the previous paper by Hawking and Turok \cite{HTnew}, contradicts the proposal to maximize the Euclidean action.

One should note, that Eq. (\ref{newaction}) is not a unique way to write the action for the theory of the field $F$. One can add to the action the integral of a total derivative
\begin{equation}\label{Towns}
S_{\rm extra} = 
{\alpha\over 24}  \int d^4 x\,  \epsilon^{\mu \nu \rho \lambda} F_{\mu \nu \rho \lambda} \ ,
\end{equation}
as proposed by Aurelia, Nicolai and Townsend \cite{ANT}. Here $\alpha$ is an arbitrary constant. Since this is a total derivative, it does not change the instanton solution, it does not modify the entropy, but it gives an extra contribution to the Euclidean action $\Delta S \sim -{\alpha\over c} {3M_p^4 \over 8 V(\phi)^2}  F^2 $. For $\alpha = c$, this term cancels the $F^2$ term in (\ref{modaction}) \cite{Duff}. Thus, depending on $\alpha$ one gets different expressions for the Euclidean action, whereas the expression for the entropy is $\alpha$-independent. This suggests that one should look for the extremum of the entropy rather than of the  action. 
 
One may try to resolve the ambiguity by applying stochastic approach. In this case the presence of the $F$ field will be entirely irrelevant as long as its contribution to the vacuum energy is cancelled by $V_0$. One obtains the same stationary probability distribution  (\ref{E38a}) determined by the exponent of the entropy $e^{\rm \bf S}$, independently of the existence of the field $F$.  This means that the presence of the field $F$  cannot change the prediction $\Omega = 0.01$ based on the use of the Hartle-Hawking wave function.  

In a new version of their paper \cite{HT2} Hawking and Turok agreed with our conclusion. They noted that if one properly takes into account all boundary terms, an expression for the Euclidean action changes, and the disagreement between the calculation using the action and the entropy disappears \cite{Dunkan,HT2}. This implies, just as we argued above,  that the introduction of the field $F$ in this model does not resolve the problem of having too small value of $\Omega$.

A potentially interesting consequence of the introduction of the $F$-field   is the cosmological constant problem. In order to analyse it, in the new version of their paper \cite{HT2} Hawking and Turok reverted the sign of the $F^2$ term in the action, to bring it closer to the Freund-Rubin work on supergravity compactification \cite{FR}. 
The exponent of the entropy $e^{\rm \bf S}$ can be represented as
\begin{equation}\label{entropy2}
P \sim e^{\rm \bf S} =
\exp\left({3M_p^4\over 8 (V(\phi)+V_0 + \rho_F)}\right) ,
\end{equation}   
where $\rho_F$ is the (negative) energy density of the $F$-field. 
  \cite{HAWK}. If one interprets this result as the probability of the quantum creation of the universe, this may imply that the universe should be created in a state corresponding to the minimal value of the total energy density $V(\phi)+V_0 + \rho_F \ll M_p^4$ consistent with the subsequent emergence of life. The possibility of creation of  universes with different $\rho_F$ then allows us to use anthropic principle to make the observable value of the cosmological constant very small \cite{HT2}.
However, in this case one still has the problem of living in a structureless universe with $\Omega = 0.01$.

On the other hand, if one uses the tunneling wave function, one finds
\begin{equation}\label{tunncosm}
P \sim  
\exp\left(-{3M_p^4\over 8 (V(\phi)+V_0 + \rho_F)}\right) ,
\end{equation}   
This implies that the universe is created in a state with $V(\phi)+V_0 + \rho_F \sim M_p^4$. Note that the distribution of probability of creation of a universe in this scenario is practically flat with respect to $\rho_F$ in an enormously wide interval $\Delta  \rho_F \sim M_p^4$. Thus anthropic principle easily fixes  $|V_0 + \rho_F| \lesssim 10^{-29}$ $g/cm^3$, which solves the cosmological constant problem. The  initial value of $V(\phi)$ in this scenario is $O(M_p^4)$, which leads to a very long stage of inflation with $\Omega = 1$, or to $\Omega < 1$ in the models introduced in \cite{Gott,BGT,Open}.

One should note, however, that the possibility to resolve the cosmological constant problem in realistic theories involving the field $F$ requires additional investigation. Indeed, the value of the (negative) energy density of this field in the models based on supergravity depends on the radius of compactification. In realistic models one  expects $\rho_F \sim - M_p^4$. If, depending on compactification,   $\rho_F$ may take only a discrete set of values such that $\rho_F \sim - M_p^4$, the solution of the cosmological constant problem in this scenario would require that $V_0$ coincides with one of these values with an accuracy  $10^{-123}M_p^4$. Thus   the introduction of the antisymmetric tensor field $F$  does not help to solve the problem of having too small $\Omega$ in the model of \cite{HT2}, and the possibility that it can help us to solve the cosmological constant problem also remains rather problematic.

\section{Conclusions}

Prior to the invention of the inflationary universe scenario it seemed that
quantum cosmology  is very important for
understanding  the underlying principles of the theory of evolution of the
universe, but it may not have any observational consequences.   During the last
15 years quantum cosmology has become
a more established science, which allows us
to
make testable observational predictions.

As we have seen, both the Hartle-Hawking and the tunneling wave function of the
universe can describe creation of an open inflationary universe. This is a very
interesting possibility in view of the recent tendency to claim that the
observations favor smaller value of $\Omega$.

However, different versions of quantum cosmology predict completely different
values of $\Omega$. The Hartle-Hawking wave function predicts that if the
universe is closed, then $\Omega \gg 1$, and if it is open, one has $\Omega
\sim
10^{-2}$. This is experimentally unacceptable.  In this paper we confirmed that
this result is practically model-independent if galaxy formation occurs due to
adiabatic density perturbations produced during inflation.   One may try to
avoid this conclusion by appealing to some unspecified versions of string
theory or M-theory where the situation might be better \cite{HT,HTnew}. But in
the absence of any realization of this idea one may conclude that at the
present time the Hartle-Hawking wave function, {\it if used to calculate the
probability of quantum creation of the universe}, is in a direct contradiction
with observational data.

Is it really possible to rule out the Hartle-Hawking wave function on the basis
of these results? Perhaps such a conclusion would be premature. The main
argument which pushed the most probable value of $\Omega$ toward $10^{-2}$ was
based on the equation for adiabatic density perturbations in a theory of a
single
scalar field, Eq. (\ref{PROB2}). This conclusion can  change  if adiabatic
perturbations are very small, and perturbations responsible for galaxy
formation are isocurvature, or  if  they are produced by topological defects.
For example, one may imagine that the phase transition which leads to the
formation
of  topological defects occurs during the last stages of chaotic inflation, see
e.g. \cite{KL}. Then the defect production is a threshold effect, which occurs
only if the universe is formed with a sufficiently large scalar field $\phi$.
In such a situation the Hartle-Hawking wave function will suggest that the
scalar field should be as small as possible, but still large enough for the
phase transition to take place, because  density perturbations would be too
small in the universe without strings. Then the unfortunate prediction $\Omega
= 10^{-2}$ may disappear, but it will be replaced by the fine-tuning of the
moment of onset of
the phase transition. Also, the possibility to produce the large scale
structure of the universe using isothermal perturbations or topological defects
is currently out of favor, so we are not sure whether one should consider it
seriously. 

In our opinion, the whole problem appears here because one tries to
apply the Hartle-Hawking wave function for the investigation of the probability
of creation of the universe. Our analysis of this issue contained in Sections
II and III suggests that it   should not be used for that purpose. In particular, we have seen that stochastic approach to inflation unambiguously produces the same probability distribution as the Hartle-Hawking wave function, see Eq. (\ref{E38a}). This equation
has a simple interpretation: the Hartle-Hawking wave function (in agreement with its derivation in \cite{HH}) describes the probability distribution to find the field $\phi$ in a stationary state (if this state exists) {\it after} the field relaxes towards the minimum of the effective potential. This wave function does not describe creation of the universe, inflation and the process of relaxation toward this ground state, which is the main subject of our investigation.

If one
uses the tunneling wave function for the description of initial conditions in
the
universe,   then in most inflationary models the universe should have $\Omega =
1$, which agrees with the standard expectation that inflation makes the
universe flat. This result is not  sensitive at all to the exact features of the tunneling wave function, and in fact to the very use of the tunneling wave function. The only thing which we need to assume is that  there is no exponential suppression of quantum creation of a very small universe  as compared to the probability of creation of a very large universe \cite{book}.

Moreover,  according to the theory of
a self-reproducing inflationary universe, which applies to most versions of chaotic inflation \cite{b19}, one can avoid making even this assumption. The theory of a self-reproducing universe  asserts that initial conditions
are nearly irrelevant for the description of the properties of the main part of
the universe \cite{Mijic,LLM}. In most models of that type one has $\Omega = 1$ after inflation.

There
exists a new potentially interesting class of models where creation of an open
universe described by the tunneling wave function may be possible. A thorough
investigation is needed in order to verify   whether this possibility is
realistic or
not. There are many reasons to be sceptical about it, see Sect. \ref{tunn} and also \cite{VIL,BL}. It is important, however, that independently of this
possibility we still
have the class of models proposed in \cite{Gott,BGT,Open}, which does not seem
to   work  in the context of the Hartle-Hawking proposal, but which is quite
compatible with the tunneling wave function of the universe, as well as with
the theory of a self-reproducing inflationary universe.

Investigation of quantum cosmology in application to the open universe creation
is very difficult. Much work is to be done in order to investigate the new
possibilities which we now have. However, one should not underestimate the
recent progress. Until very recently, we did not have {\it any} consistent
cosmological models describing a homogeneous open universe. Even though the
open universe model did exist from the point of view of mathematics, it simply
did not appear to
make any sense to assume that all parts of an infinite universe can be
created simultaneously  and have the same value of energy density everywhere.

That is why it is very encouraging  that during the last few years we have
found several different mechanisms of creation of an open
universe. All of these mechanisms require the universe to be inflationary.  It
is still true that inflationary models describing the universe with $\Omega =
1$ are much simpler than the models with $\Omega \not = 1$. Hopefully, the
universe will appear to be flat, and we will never need to use any of the
models of open inflation. But if we find out that Nature has chosen to build
the universe in a way which does not look
particularly natural, this may give us a rare opportunity to reexamine some of
our ideas  and  to learn more about quantum  cosmology.

\subsection*{Acknowledgments}

It is a pleasure to thank  J. Garc\'{\i}a--Bellido and  R. Kallosh for useful
discussions. I am especially grateful to R. Bousso,   N. Kaloper, and L.A. Kofman  for valuable discussions and suggestions at different
stages of this
investigation. This work   was supported in part by NSF
Grant  No. PHY-9219345.

\end{document}